\documentclass{article}
\pdfoutput=1
\usepackage{arxiv}

\usepackage{authblk}
\usepackage[utf8]{inputenc} 
\usepackage[T1]{fontenc}    
\usepackage{hyperref}       
\usepackage{url}            
\usepackage{booktabs}       
\usepackage{amsfonts}       
\usepackage{nicefrac}       
\usepackage{microtype}      
\usepackage{lipsum}         
\usepackage{graphicx}
\usepackage{doi}

\usepackage[caption=false]{subfig}
\usepackage[linesnumbered, vlined, ruled]{algorithm2e}
\usepackage{amsmath}
\usepackage{cleveref}       
\usepackage[export]{adjustbox}

\newtheorem{theorem}{Definition}

\title{Accelerating Spatio-Textual Queries with Learned Indices}

\begin{document}

\author[1]{Georgios Chatzigeorgakidis}
\author[2]{Kostas Patroumpas}
\author[3]{Dimitrios Skoutas}
\author[4]{Spiros Athanasiou}

\affil[1]{IMSI, Athena R.C., Greece, \textit{email: gchatzi@athenarc.gr}}
\affil[2]{IMSI, Athena R.C., Greece, \textit{email: kpatro@athenarc.gr}}
\affil[3]{IMSI, Athena R.C., Greece, \textit{email: dskoutas@athenarc.gr}}
\affil[4]{IMSI, Athena R.C., Greece, \textit{email: spathan@athenarc.gr}}

\maketitle

\begin{abstract}
    Efficiently computing spatio-textual queries has become increasingly important in various applications that need to quickly retrieve geolocated entities associated with textual information, such as in location-based services and social networks.
To accelerate such queries, several works have proposed combining spatial and textual indices into hybrid index structures. Recently, the novel idea of replacing traditional indices with ML models has attracted a lot of attention. This includes works on learned spatial indices, where the main challenge is to address the lack of a total ordering among objects in a multidimensional space. In this work, we investigate how to extend this novel type of index design to the case of spatio-textual data. We study different design choices, based on either loose or tight coupling between the spatial and textual part, as well as a hybrid index that combines a traditional and a learned component. We also perform an experimental evaluation using several real-world datasets to assess the potential benefits of using a learned index for evaluating spatio-textual queries.

\end{abstract}

\section{Introduction}
\label{sec:intro}

The ubiquity of mobile devices, along with the proliferation of location-based services and geosocial networks, has set off an unprecedented generation of geotagged content. Efficiently querying and exploring spatio-textual data has become increasingly important for various applications. Typically, query execution must be very fast in order to enable interactive response times. To address this need, several spatio-textual indices have been proposed over the past decade. These are hybrid structures that combine a spatial index, such as grid or R-tree, with a textual part, such as bitmap or inverted file, to allow pruning the search space using both criteria.

Recently, the novel idea of thinking of indices as ML models has been proposed~\cite{kraska2018case}. According to it, an index can be formulated as a function $f$ that maps an input key to a set of candidate objects (e.g., stored in a page on disk). Naturally, this has attracted a lot of interest, as it opens up a fundamentally new approach to the decades old problem of index design. Initial works focused on designing learned indices for one-dimensional data, leveraging the total ordering that exists among such objects.

    
Following this novel line of research, {\em learned spatial indices} have been recently introduced~\cite{davitkova2020ml, qi2020effectively}, extending the concept of learned indices from 1-dimensional data to two or more dimensions. Coping with more than one dimension raises additional challenges, most importantly the lack of a total ordering among objects. This is typically addressed by employing a space-filling curve (SFC) to impose an ordering. However, the higher the dimensionality, the less the accuracy of the resulting ordering, which also means a higher error for the ML model that tries to learn the data distribution based on it. Consequently, learned indices over multidimensional data typically focus on a low number of dimensions.

Many spatial objects, such as Points of Interest or geotagged posts in social media, are described by some textual information, which is or can be represented as a {\em set of keywords}. Spatio-textual queries allow retrieving such objects based on both spatial and textual criteria. Perhaps the two most common types of such queries are \emph{Boolean Window Queries} (BWQ) and \emph{Boolean $k$-nearest neighbor Queries} (B$k$Q). A BWQ retrieves objects located within a rectangular spatial region that match all query keywords. A B$k$Q retrieves the $k$ nearest objects to the query location that also match all given keywords.

\begin{figure}[!t]
\centering
\setcounter{subfigure}{0}
\subfloat[A Boolean Window Query.]{\includegraphics[width=0.3\textwidth]{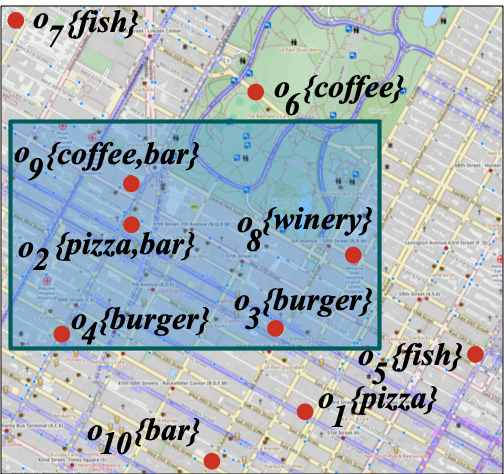}\label{subfig:motiv_window}}
\hspace{10pt}
\subfloat[A Boolean $k$NN Query.]{\includegraphics[width=0.3\textwidth]{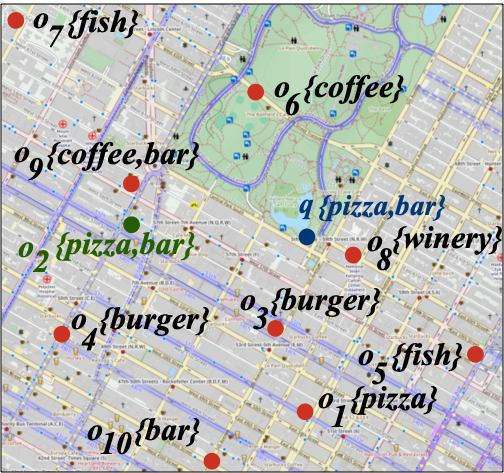}\label{subfig:motiv_knn}}
\caption{Boolean spatio-textual queries.}
\label{fig:motivation}
\end{figure}

Figure~\ref{fig:motivation} illustrates an example involving ten spatio-textual objects. Each object $o_i$ is associated with a location and a set of keywords. Figure~\ref{subfig:motiv_window} shows a BWQ $Q$ = ($\mathcal{W}$, $\langle pizza, bar \rangle$) to fetch all objects inside a rectangular region $\mathcal{W}$ (shown in blue) that contain the keywords \emph{pizza} and \emph{bar}. Figure~\ref{subfig:motiv_knn} shows a B$k$Q $Q$ = ($q$, $\langle pizza, bar \rangle$, 1), which seeks the closest object to location $q$ tagged with \emph{pizza} and \emph{bar}. In both queries, the result is object $o_2$. 

To accelerate such queries, several {\em spatio-textual indices} have been proposed over the past decade. These are hybrid structures that combine a spatial access method (e.g., grid, R-tree, quadtree, k-d tree) with a textual one (e.g., bitmap, inverted file). Proposed variants differ on the type of spatial and textual index they are based on, the type of coupling (i.e., loose {\em vs} tight), and the type of queries they support. A typical example is the IR$^2$-Tree~\cite{defelipe2008icde}, which extends the widely used spatial index R$^*$-Tree~\cite{beckmann1990r} by augmenting its nodes with textual information in the form of signatures~\cite{faloutsos1984signature}.

A question that naturally arises is whether and how the evaluation of spatio-textual queries can benefit from the use of ML models, as has been suggested for one-dimensional or two-dimensional data. In this work, we investigate this question by designing and evaluating hybrid spatio-textual indices that employ a learned component. Although some suggestions for learned indices over strings have been proposed, their benefits are not clear~\cite{kraska2018case}. Thus, for the learned part, we focus on the spatial component, in particular using the current state-of-the-art learned spatial index RSMI~\cite{davitkova2020ml}. RSMI has been shown to be able to speed up spatial queries by an order of magnitude compared to traditional query processing based on R-trees. Motivated by this, we investigate how to extend RSMI towards a {\em learned hybrid spatio-textual index}, and we study the implications of the learned part and the textual part to the performance of the resulting index structure. To the best of our knowledge, this is the first work to employ learned indices over spatio-textual data.


The contributions of our work can be summarized as follows:
\begin{itemize}
    \item We derive a baseline learned spatio-textual index by loosely coupling RSMI with an inverted index to enable textual pruning on the leaf nodes.
    \item We suggest a tightly coupled index that augments RSMI by associating both internal and leaf models with textual information in the form of bitmaps.
    \item We design a hybrid index that combines both a learned part (RSMI) and a traditional part (IR$^2$-Tree).
    \item We present an alternative partitioning strategy that takes into account the textual information besides the spatial one.
    \item We empirically evaluate the performance of proposed learned index variants against real-world datasets and also compare them with traditional indices to identify potential benefits.
\end{itemize}

The remainder of this paper is organized as follows: Section~\ref{sec:background} reviews related work.  
In Section~\ref{sec:problem} we formulate the problem. The methods under examination are presented in Section~\ref{sec:spatext_learned}, followed by an extensive experimental evaluation in Section~\ref{sec:exp}. Finally, Section~\ref{sec:conclusions} concludes the paper.

\section{Background and Related Work}
\label{sec:background}

In this section, we briefly review related work on spatio-textual indices and learned indices.



\vspace{1mm}
\noindent \textbf{Spatio-Textual Indices.}
Spatio-textual indices are hybrid index structures that combine a spatial and a textual component so that pruning on both criteria can be performed during query processing. Several such indices have been proposed in the literature (see surveys in~\cite{christoforaki2011cikm,chen2013pvldb,DBLP:journals/vldb/ChenCCJ21}). The spatial indexing part can be based on R-trees (e.g., IF-R$^*$-Tree and R$^*$-Tree-IF~\cite{DBLP:conf/cikm/ZhouXWGM05}, KR$^*$-Tree~\cite{DBLP:conf/ssdbm/HariharanHLM07}, IR$^2$-Tree~\cite{defelipe2008icde}, IR-Tree~\cite{DBLP:journals/vldb/WuCJ12}, SKI~\cite{cary2010ssdbm}, S2I~\cite{rocha2011ssd}, WIBR-Tree~\cite{DBLP:journals/tkde/WuYCJ12}), Quad-tree (e.g., I$^3$~\cite{DBLP:conf/edbt/ZhangTT13}, ILQ~\cite{DBLP:journals/tkde/ZhangZZL16}), grid (e.g., ST and TS~\cite{DBLP:conf/ssd/VaidJJS05}, SKIF~\cite{DBLP:conf/dexa/KhodaeiSL10}) or Space-filling curve (e.g., SF2I~\cite{chen2006sigmod}, SFC-QUAD~\cite{christoforaki2011cikm}). For the textual part, typically a signature (e.g., a bitmap) or an inverted file is used to indicate the presence of keywords in the objects being indexed. In our approach, we employ the IR$^2$-tree~\cite{defelipe2008icde}, which is a representative spatio-textual index that can answer both Boolean window (BWQ) and $k$NN (B$k$Q) queries. It is based on the R$^*$-Tree~\cite{beckmann1990r}, where each node is enhanced with a signature indicating which keywords exist in the objects corresponding to that node.

\vspace{1mm}
\noindent \textbf{Learned Indices.}
Recently, the novel idea of learned indices has been proposed~\cite{kraska2018case}. Such indices rely on ML models to predict the physical location of an object in memory or on disk. The key idea is to use a machine learning model, which may range from a deep neural network to a simple regression model, to learn the cumulative distribution function (CDF) of object keys in a sorted array. Thus, most of these indices are proposed for 1-dimensional data, where a natural ordering of the objects exists~\cite{DBLP:journals/pvldb/MarcusKRSMK0K20}.

Nevertheless, several multidimensional learned indices have also been proposed to handle data of two or more dimensions, overcoming the lack of an inherent total ordering~\cite{DBLP:conf/gis/Al-MamunWA20,DBLP:conf/sigmod/NathanDAK20,ding2020tsunami}. These often rely on a space-filling curve to map the data to a single dimension, so that a total ordering can be imposed. For instance, the $z$-order model (ZM) index~\cite{wang2019learned} combines the $z$-order space-filling curve with a model to learn the distribution of the data. LISA~\cite{li2020lisa} is a learned index structure for spatial data, which consists of a function that maps spatial keys into 1-D values, a function that partitions the mapped space, and a series of local models that organize partitions into pages. In this paper, we use the Recursive Spatial Model Index (RSMI)~\cite{qi2020effectively} as the basis for designing a learned spatio-textual index, hence we describe RSMI in more detail below.

RSMI~\cite{qi2020effectively} relies on a $z$-order space-filling curve to map the 2-D spatial points to a single dimension, where the resulting CDF of the data is learned. To avoid large gaps in this CDF, it first transforms the original point coordinates to a rank space, following the same idea that has been proposed as an R-Tree packing strategy~\cite{qi2018theoretically}. Hence, the $z$-order curve is applied to the rank space instead of the original one. Then, RSMI employs a feed forward neural network to learn the resulting CDF. The maximum error $\epsilon$ of the model is also calculated and is used at query time to define a search range around the predicted location so that results are not missed. 

When indexing a large collection of spatial objects, it is often not feasible for a single model to learn the entire CDF with sufficient accuracy. To overcome this, RSMI recursively partitions the space and creates a hierarchy of models, so that each model only learns a fraction of the entire CDF. The partitioning is a non-regular grid that follows the data distribution, assuming a maximum allowed partition size $S$. At each level of the hierarchy, the partitions are ordered based on a $z$-order curve calculated over their centroids. For each partition, a separate model is trained. At query time, each model predicts which of its children nodes should be visited next, thus traversing the hierarchy from the root to the leaf nodes. At the bottom level, each leaf model predicts a range of blocks to be searched, also taking into account the model error $\epsilon$.

\section{Problem Definition}
\label{sec:problem}

In this section, we introduce the main definitions and notions used throughout the paper. The basic notations are listed in Table~\ref{tab:notations}.



Assume a collection $\mathcal{D}$ of spatio-textual objects, which are formally defined as follows.

\begin{theorem}[Spatio-Textual Object] A spatio-textual object $o$ is represented by a pair $(o.s, o.T)$, where $o.s$ is a point in a 2-dimensional Euclidean space and $o.T$ is a set of keywords.
\end{theorem}

We consider the following two basic types of spatio-textual queries, which are commonly studied in the literature~\cite{christoforaki2011cikm,chen2013pvldb,DBLP:journals/vldb/ChenCCJ21}.





\begin{theorem}[Boolean Window Query (BWQ)] A boolean window spatio-textual query is defined as $BWQ = (\mathcal{W}, \mathcal{T})$, where $\mathcal{W}$ is a rectangular spatial area (window), and $\mathcal{T}$ is a set of keywords. Given a collection $\mathcal{D}$ of spatio-textual objects, the result set $R$ of $BWQ$ comprises all objects that lie within $\mathcal{W}$ and contain all the keywords in $\mathcal{T}$, i.e., $R = \{ o \in \mathcal{D} \, | \, within(o.s, \mathcal{W}) \wedge \mathcal{T} \subseteq o.T \}$.
\end{theorem}

If the sequel, for purposes of presentation, we interchangeably denote a spatial window $\mathcal{W}$ with the pair $(q_l, q_h)$ of its lower left and upper right corners.

\begin{theorem}[Boolean $k$NN Query (B$k$Q)] A boolean $k$NN spatio-textual query is defined as $BkQ=(q, \mathcal{T}, k)$, where $q$ is a point location, $\mathcal{T}$ is a set of keywords, and $k$ is the number of objects to be returned. Given a collection $\mathcal{D}$ of spatio-textual objects, the result set $R$ of $BkQ$ comprises $k$ objects that contain all the keywords in $\mathcal{T}$ and are most closely located to $q$, i.e., (i) $R \subseteq \mathcal{D}$, (ii) $|R| = k$, (iii) $\forall o \in R, \mathcal{T} \subseteq o.T$, (iv) $\forall o \not \in R, \exists o' \in R \, | \, d(q. o.s) \geq d(q, o'.s)$, where $d$ refers to Euclidean distance.
\end{theorem}



Our goal in this paper is to design learned spatio-textual indices that can accelerate the evaluation of $BWQ$ and $BkQ$ queries compared to traditional indexing schemes.


\begin{table}[t]
\centering
\caption{Basic notations}
\begin{tabular}{cl}
\hline
$\mathcal{D}$ & dataset of spatio-textual objects \\
$o.s$ &  location of spatio-textual object $o \in \mathcal{D}$ \\
$o.T$ & set of keywords associated to spatio-textual object $o \in \mathcal{D}$ \\
$\mathcal{T}$ & set of query keywords \\
$\mathcal{W}$ & spatial query rectangle (window) in BWQ \\
$q_l, q_h$ & lower left and upper right corner of window $\mathcal{W}$ in BWQ \\
$mbr$ & Minimum Bounding Rectangle over the input location(s) \\
$k$ & number of nearest qualifying objects to retrieve in B$k$Q \\
$\ell$ & side length of window used in evaluating B$k$Q \\
$R$ & set of query results \\
$S$ & maximum size (number of contained objects) per partition  \\
$B$ & block of data in a leaf node of RSMI \\
$\epsilon$ & maximum error of the model \\
\hline
\end{tabular}
\label{tab:notations}
\end{table}

\section{Index Design}
\label{sec:spatext_learned}

Inspired by traditional spatio-textual indices, which combine a spatial part and a textual part, our approach is to enhance a learned spatial index with keyword information.
As a basis, we use RSMI~\cite{qi2020effectively} (see Section~\ref{sec:background}). We first provide an overview of our approach, and then we describe three indices based on different design choices, namely a loosely coupled, a tightly coupled, and a hybrid scheme.

\subsection{Overview}
\label{sec:overview}


\SetAlFnt{\small}
\begin{algorithm}[!t]
	\DontPrintSemicolon

    \SetKwProg{BWQTraversal}{Procedure}{}{}
    \BWQTraversal{\texttt{BWQTraversal}$(N, Q, R)$}{
    
        $M \leftarrow N.model$
        
        \If{$N$ is $leaf$}{
            \ForEach{$B \in [M(Q.q_l) - M.\epsilon_l, M(Q.q_h) + M.\epsilon_h]$}{
                \If{$intersects(B.mbr, Q.\mathcal{W})$}{
                    $R \leftarrow CheckLeafNode(B, Q, R)$
                }
            }
        }
        \Else{
            \ForEach{node $N' \in [M(Q.q_l),M(Q.q_h)]$}{
                \If{$CheckInnerNode(N', Q)$}{
                    \texttt{BWQTraversal}$(N', Q, R)$ \\
                }
            }
        }
        \KwRet $R$
    }

\caption{\texttt{Index Traversal}}
	\label{alg:traversal}	
\end{algorithm}

\begin{figure*}[!t]
    \centering    
    \includegraphics[width=0.85\linewidth]{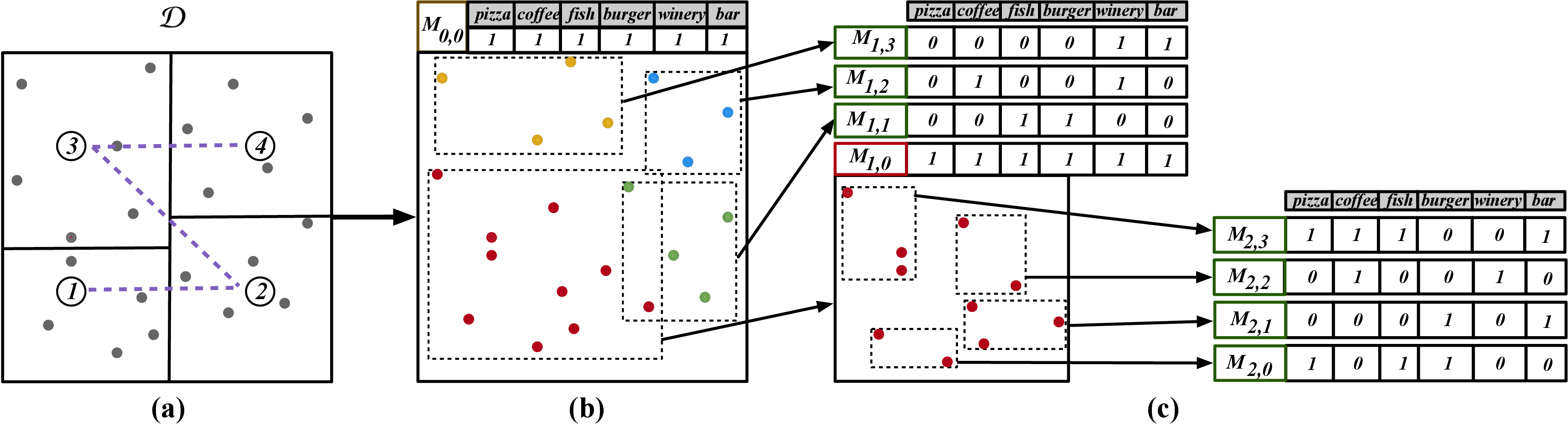}
    \vspace{-7pt}    \caption{The RSMI-BM spatio-textual index.}
    \label{fig:rsmi_st}
\end{figure*}

The backbone of our proposed indices is RSMI, which hierarchically partitions the space and builds a respective hierarchy of ML models. At each level, partitions are created. Each partition is assigned an id based on $z$-order, and is associated with a model that directs the search to its children. An illustrative example is depicted in Figure~\ref{fig:rsmi_st} (the bitmaps next to each model $M_{i, j}$ are explained later).
There are four partitions at the top level. The root model is $M_{0,0}$, which points to the four models $M_{1,0}, ..., M_{1,3}$ at the next level. Models $M_{2,0}, ..., M_{2,3}$ are leaf models.
Each model is also associated with the MBR of the respective partition. This is used during query evaluation to avoid false positives that may occur due to prediction errors by the models, i.e., to prune subtrees whose MBR does not intersect with the specified query window.

To answer a {\em Boolean Window Query} $Q$, we traverse the index as in Algorithm~\ref{alg:traversal}. Procedure \emph{BWQTraversal} is recursive, starting from the root node. Assume the model $M$ of a visited node $N$. Given the lower left and upper right corners ($q_l$ , $q_h$) of the query window $\mathcal{W}$, model $M$ predicts the range of the children nodes (or blocks, in the case of leaf models) across the $z$-order curve that should be checked. 
For inner nodes (Lines 7-10),  process \emph{CheckInnerNode} takes into account any available information (MBR, bitmaps) per node and determines whether to recursively descend its subtree, as detailed later for each of the proposed schemes. To reduce the number of total checks, note that RSMI does not consider the model errors at inner nodes. 
For each block in a visited leaf node (Lines 3-6), if its MBR intersects the query window, we proceed to check its contents according to the keyword indexing scheme applied by each of our proposed methods. At the leaf level, the model errors $\epsilon_l, \epsilon_h$ are taken into account.  
The process terminates once there are no more leaves to search and the set $R$ of results is returned.

To answer a {\em Boolean $k$NN Query} (B$k$Q), we follow the same approach as in RSMI. This involves executing a series of BWQ with increasingly larger window size until at least $k$ objects are found. As a heuristic, the side length of the initial window is set to $\ell=\sqrt{k/|\mathcal{D}|}$ (assuming that original coordinates have been normalized to a unit space). Then, this length is increased by a factor of 2 at each subsequent window query. In all proposed indexing schemes, the applied query is a $BWQ$ query, i.e., only objects containing the query keywords are qualifying for the top-$k$ results.



\subsection{Loosely Coupled Scheme}
\label{subsec:loosely}


This approach loosely combines a spatial and a textual indexing component. Its advantage is simplicity and modularity, allowing existing implementations to be reused. The downside is lower pruning capacity, since each component only allows pruning with the respective criterion.

\vspace{1mm}
\noindent \textbf{Index Structure.}
In this scheme, we combine RSMI with inverted files (IF) attached to its leaf nodes. We refer to the resulting index as {\em RSMI-IF}. Given a collection $\mathcal{D}$ of spatio-textual objects, RSMI is first built based on their point locations. Recall that each leaf model points to a series of blocks. RSMI-IF constructs an inverted index for each such block. Hence, at query time, instead of checking all objects in each block, only those that contain a query keyword can be accessed. Essentially, this is analogous to the R$^*$-Tree-IF~\cite{DBLP:conf/cikm/ZhouXWGM05} spatio-textual index, where we use the learned spatial index RSMI in place of the R$^*$-Tree.



\vspace{1mm}
\noindent \textbf{Query Processing.}
Following the loosely-coupled nature of the RSMI-IF index, evaluation of a Boolean window query $BWQ = (\mathcal{W}, \mathcal{T})$ comprises two main stages. 
The first stage is essentially a spatial window query over RSMI using the specified window $\mathcal{W}$. Inner nodes with MBRs not intersecting $\mathcal{W}$ can be safely pruned (Algorithm~\ref{alg:loosely}, process \emph{CheckInnerNode}). However, instead of returning all objects within $\mathcal{W}$, the process terminates a step earlier, specifically as soon as the leaf models identify the candidate blocks where these points may be contained. The second stage is handled by process \emph{CheckLeafNode} in Algorithm~\ref{alg:loosely}. For each such block in a leaf, its associated inverted file $IF$ is used to retrieve the inverted lists of objects containing the query keywords. The intersection of these lists produces a set of candidate objects that contain all the query keywords. Finally, the location of each candidate object $o$ is checked to verify that it is within $\mathcal{W}$. This processing method is also employed to answer B$k$Q queries, as noted in Section~\ref{sec:overview}.

\SetAlFnt{\small}
\begin{algorithm}[!t]
	\DontPrintSemicolon

    \SetKwProg{CheckInnerNode}{Procedure}{}{}
    \CheckInnerNode{\texttt{CheckInnerNode}$(N, Q)$}{
        \lIf{intersects($N.mbr, Q.\mathcal{W}$)}{
            \KwRet true
        }
        \lElse{
            \KwRet false
        }
    }
    
    \vspace{4pt}    

    \SetKwProg{CheckLeafNode}{Procedure}{}{}
    \CheckLeafNode{\texttt{CheckLeafNode}$(B, Q, R)$}{
        $cands \leftarrow search(B.IF, Q.\mathcal{T})$ \\
        \ForEach{object $o \in cands$}{
            \lIf{within($o.s, Q.\mathcal{W}$)}{
                    $R \leftarrow R \cup \{o\}$
            }
        }
        \KwRet $R$
    }

\caption{\texttt{RSMI-IF query processing}}
	\label{alg:loosely}	
\end{algorithm}

\subsection{Tightly Coupled Scheme}
\label{subsec:tightly}

In this setting, the proposed index tightly combines a spatial and a textual component, so that pruning with both criteria can be performed throughout all levels of the hierarchy. Next, we introduce {\em RSMI-BM}, an index that augments RSMI with the use of bitmaps to encode the presence of keywords in the objects.


\vspace{1mm}
\noindent \textbf{Index Structure.}
RSMI-BM follows the same base structure as RSMI, i.e., it consists of a hierarchy of models. However, each model is now additionally associated with a bitmap, where each bit corresponds to a keyword. The bit corresponding to keyword $t$ is set to 1, if there exists an object in the subtree of that model that contains $t$, or 0 otherwise. To construct the index, the RSMI index is  built first, based on the locations of the objects. Then, its nodes are traversed bottom-up, keeping track of the encountered keywords, which are used to set the respective bitmap for each model. An RSMI-BM index with three levels is illustrated in Figure~\ref{fig:rsmi_st}, corresponding to the collection of spatio-textual objects depicted in Figure~\ref{fig:motivation}.

\vspace{1mm}
\noindent \textbf{Query Processing.}
Evaluating a Boolean window query BWQ 
again follows the paradigm shown in Algorithm~\ref{alg:traversal}. The model hierarchy of the RSMI component is traversed from the root to the leaves, visiting the nodes indicated by each model encountered at each level. Checking the given spatial window $\mathcal{W}$ against node MBRs is also used for pruning, as in RSMI-IF. The difference now is that before using a model to make a prediction, the algorithm also checks whether the bitmap $bm$ associated to the corresponding leaf or inner node $N$ (Algorithm~\ref{alg:check_bitmap}) matches the bitmap of the query keywords. If $bm$ has a bit set to 0 for any of the query keywords, then that model is skipped, i.e., its corresponding subtree is pruned.


\begin{algorithm}[!t]
	\DontPrintSemicolon
	
    \vspace{4pt}
    \SetKwProg{CheckInnerNode}{Procedure}{}{}
    \CheckInnerNode{\texttt{CheckInnerNode}$(N, Q)$}{
        \lIf{intersects($N.mbr, Q.\mathcal{W}$) $\wedge$ \ BitmapMatch($N.bm, bm(Q.\mathcal{T})$)}{
            \KwRet true
        }
        \lElse{
            \KwRet false
        }
    }

    \vspace{4pt}
    \SetKwProg{CheckLeafNode}{Procedure}{}{}
    \CheckLeafNode{\texttt{CheckLeafNode}$(B, Q, R)$}{
        \ForEach{object $o \in B$}{
            \lIf{within($o.s,Q.\mathcal{W}$) $\wedge$ BitmapMatch($bm(o.T), bm(Q.\mathcal{T})$)}{
                    $R \leftarrow R \cup \{o\}$
            }
        }
        \KwRet $R$
    }

    
\caption{\texttt{RSMI-BM query processing}}
	\label{alg:check_bitmap}	
\end{algorithm}


\begin{figure*}[!t]
\centering
\setcounter{subfigure}{0}
\subfloat[Partitioning on $x$-axis.]{\includegraphics[width=0.28\textwidth, valign=t]{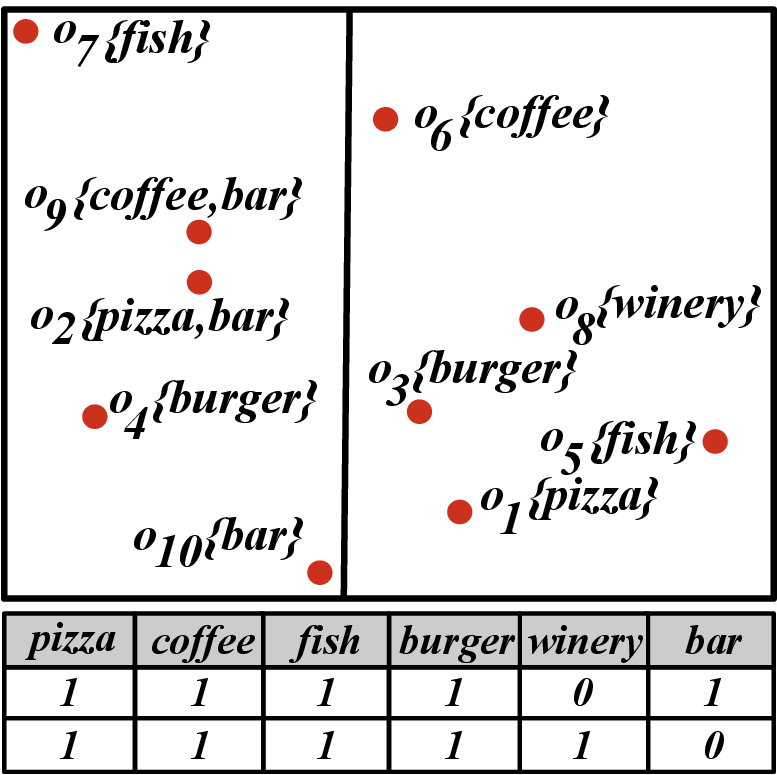}\label{subfig:alt_part1}}
\hspace{30pt}
\subfloat[Partitioning on $y$-axis.]{\includegraphics[width=0.28\textwidth, valign=t]{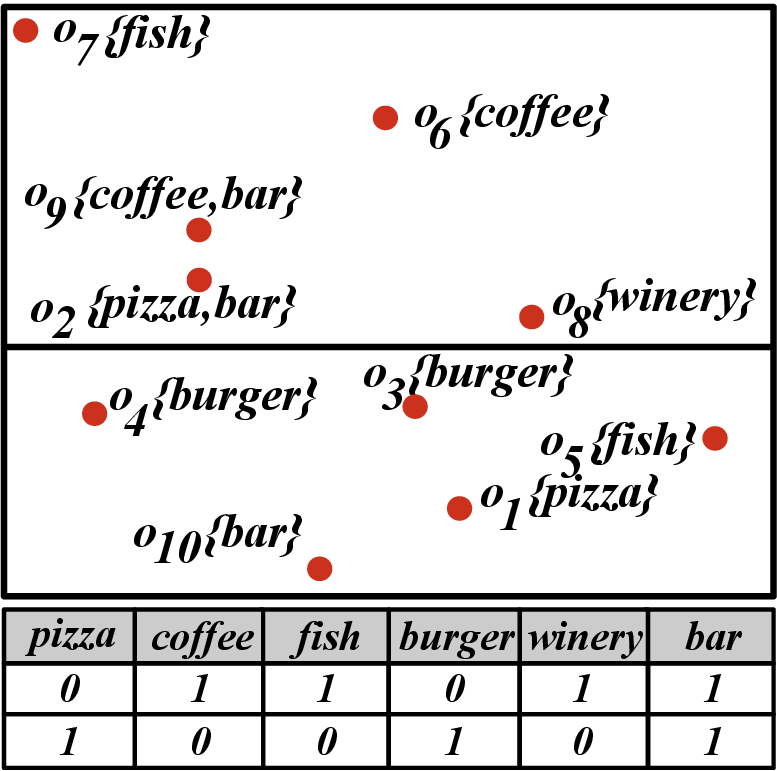}\label{subfig:alt_part2}}
\hspace{30pt}
\captionsetup[subfloat]{captionskip=30pt}
\subfloat[Final partitioning.]{\includegraphics[width=0.28\textwidth,valign=t]{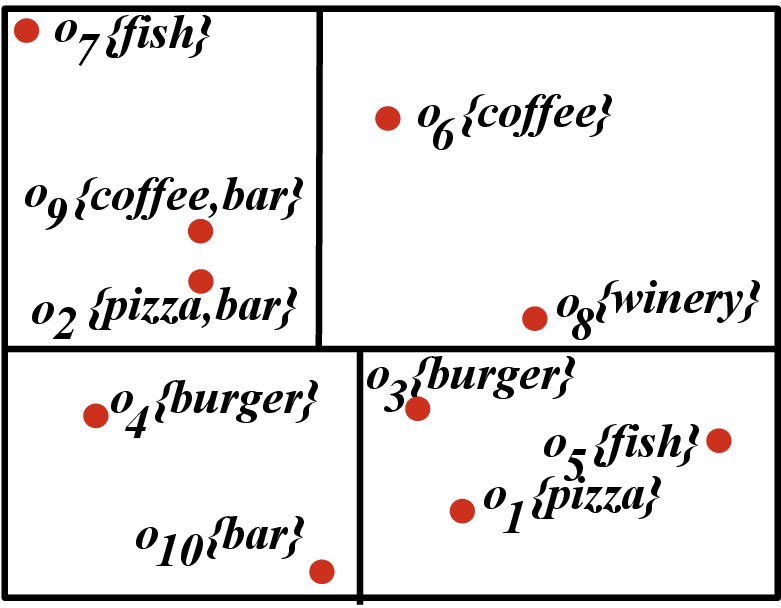}\label{subfig:alt_part3}}
\caption{The alternative partitioning method.}
\label{fig:alt_part}
\end{figure*}

\vspace{1mm}
\noindent \textbf{Alternative Partitioning.}
Recall that RSMI partitions the data based on a non-regular grid adapted to the spatial distribution of the objects. This involves interchangeably splitting the $x$ and $y$ axes into equi-sized partitions. Having balanced partitions facilitates model training. Nevertheless, RSMI completely disregards the presence of keywords in the objects. Next, we describe a modified partitioning scheme that also takes the keywords into consideration. The goal is to produce partitions with more diverse bitmaps, which will increase the pruning capacity based on the textual information. We use the Hamming distance $H$ to measure the difference between two bitmaps. Let $bm_q$ denote the bitmap corresponding to the set of query keywords $Q.\mathcal{T}$, and $bm_i$, $bm_j$ the bitmaps of two nodes $N_i$, $N_j$, respectively, at a given level of the hierarchy. Intuitively, the higher the distance $H(bm_i, bm_j)$, the less likely it is that both $bm_i$ and $bm_j$ match with $bm_q$, i.e., that both subtrees rooted at $N_i$ and $N_j$ need to be traversed. 

Based on this, our modified partitioning scheme works as follows. At each iteration, we perform two splits, one on the $x$ axis and another one on the $y$ axis, each producing two equi-sized partitions. We choose the order of these splits based on the above criterion. Let $bm^x_1$ and $bm^x_2$ (resp., $bm^y_1$ and $bm^y_2$) denote the bitmaps of the resulting partitions by splitting on the $x$ (resp., $y$) axis. If $H(bm^x_1, bm^x_2) > H(bm^y_1, bm^y_2)$, we first split on the $x$ axis, otherwise we first split on $y$. The process is repeated recursively until no partition exceeds the maximum allowed size $S$.

Figure~\ref{fig:alt_part} illustrates an example. Performing a split on the $x$ axis (Figure~\ref{subfig:alt_part1}) yields two bitmaps with Hamming distance 2. Instead, a split on the $y$ axis (Figure~\ref{subfig:alt_part2}) yields two bitmaps with Hamming distance 5. Thus, we choose to split on the $y$ axis first, resulting on the partitions shown in Figure~\ref{subfig:alt_part3}.

\subsection{Hybrid Scheme}
\label{subsec:hybrid}

In the context of learned indices, hybrid index designs have been suggested, which may employ different types of models, or even non-learned indices, at different levels of the hierarchy~\cite{kraska2018case}. Following this idea, we present the {\em RSMI-BM-IR$^2$} index, which combines a learned and a non-learned part.

\begin{figure}[!t]
    \centering
\includegraphics[width=0.55\linewidth]{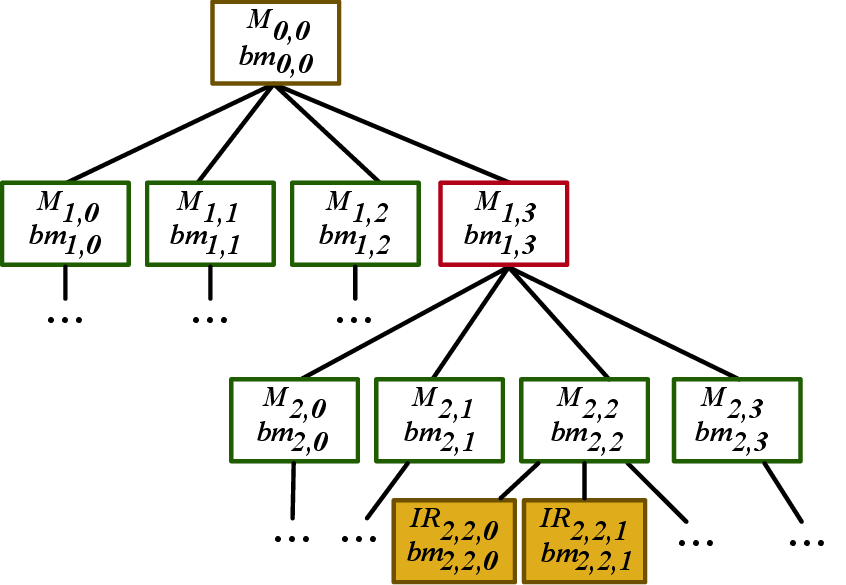}
    \caption{The hybrid RSMI-BM-IR$^2$ spatio-textual index.}
    \label{fig:hybrid}
\end{figure}

\vspace{1mm}
\noindent \textbf{Index Structure.}
In RSMI-BM-IR$^2$, the top levels in the hierarchy consist of an RSMI-BM index, as described in the previous section. The difference is that here we set a much larger maximum allowed partition size $S$, which reduces the depth of the model hierarchy and increases the number of objects contained in each leaf node. Then, for each leaf node of RSMI-BM, we index the set of its objects using an IR$^2$-tree, which is a traditional spatio-textual index (see Section~\ref{sec:background}). The structure of RSMI-BM-IR$^2$ is illustrated in Figure~\ref{fig:hybrid}. Starting from the root, there is first a hierarchy of models, with 3 levels in this example. Then, each leaf model (e.g., $M_{2,3}$) points to a series of IR$^2$-trees (i.e., $IR_{2,2,0}$, $IR_{2,2,1}$, etc.). Each of these is again associated with a respective bitmap, which is checked before starting a tree traversal.


\vspace{1mm}
\noindent \textbf{Query Processing.} Processing a BWQ using RSMI-BM-IR$^2$ proceeds similarly to the RSMI-BM until a leaf is reached. Then, as shown in Algorithm~\ref{alg:check_hybrid}, the same search query is issued against the IR$^2$-tree attached under each such leaf in RSMI-BM. The union of these search results provides the final answer.


\begin{algorithm}[!t]
	\DontPrintSemicolon
	
    \vspace{4pt}
    \SetKwProg{CheckLeafNode}{Procedure}{}{}
    \CheckLeafNode{\texttt{CheckLeafNode}$(B, Q, R)$}{
        $R \leftarrow search(B.IR^2, Q)$ \\
        \KwRet $R$
    }
    
	
\caption{\texttt{RSMI-BM-IR$^2$ query processing}}
	\label{alg:check_hybrid}	
\end{algorithm}
\section{Experimental Evaluation}
\label{sec:exp}


In this section, we present an experimental evaluation using four real-world datasets. We compare the different index variants described in Section~\ref{sec:spatext_learned}, as well as two traditional spatio-textual indices.

\subsection{Experimental Setup}
\label{subsec:exp_setup}

\subsubsection{Datasets}
\label{subsubsec:datasets}

We use four real-world datasets with different characteristics, as shown in Table~\ref{tab:datasets}. The datasets have been preprocessed to only keep frequent keywords.

\begin{table}[t]
	\centering
	\setlength{\tabcolsep}{2.75pt}
    \caption{Dataset characteristics.}
	\begin{tabular}{l|rrrr}
	{\em \textbf{Dataset}} & {\em \textbf{Objects}} & {\em \textbf{Area Size}} & {\em \textbf{Distinct kwds}} & {\em \textbf{Kwds per obj.}} \\
	\hline
	Foursquare & 94,158 & 5,053,518 $mi^2$ & 412 & 5.63 \\
    OSM USA & 1,228,616 & 3,975,720 $mi^2$ & 204 & 1.49 \\
    OSM Europe & 3,451,588 & 6,683,510 $mi^2$ & 255 & 1.71 \\
    Twitter & 10,000,000 & 6,301,967 $mi^2$ & 313 & 2.19 \\
	\end{tabular}
	\label{tab:datasets}
\end{table}

\noindent \emph{\textbf{Foursquare}}. This contains 94,158 Points of Interest (POIs)
extracted from Foursquare in the United States\footnote{\url{https://star.cs.ucr.edu/?yin/foursquare\#center=39.00,-98.00\&zoom=4}}, with a total of 412 distinct keywords and an average of 5.63 keywords per object.

\noindent \emph{\textbf{OSM USA}}. This contains 1,228,616 spatial entities extracted from OpenStreetMap (OSM) across the United States\footnote{ \url{http://download.slipo.eu/results/osm-to-csv/all/north-america/us/}}. Each entity is represented by its centroid (lon/lat coordinates). There are 204 distinct keywords, with an average of 1.49 keywords per object. 

\noindent \emph{\textbf{OSM Europe}}. This dataset contains 3,451,588 spatial entities extracted from OSM across Europe\footnote{\url{http://download.slipo.eu/results/osm-to-csv/all/europe/}}. Each entity is again represented by its centroid (lon/lat coordinates). There are 255 distinct keywords, with an average of 1.71 keywords per entity.


\noindent \emph{\textbf{Twitter}}. This contains 10 million geo-tagged tweets in the United States\footnote{\url{https://star.cs.ucr.edu/?Tweets\#center=39.00,-98.00\&zoom=4}}. There are 313 distinct keywords, with an average of 2.19 keywords per tweet.


\subsubsection{Competitors}
\label{subsubsec:competitors}

We compare the indexing schemes listed below.

\noindent \emph{\textbf{R$^*$-Tree-IF}}. A traditional spatio-textual index ~\cite{DBLP:conf/cikm/ZhouXWGM05} that loosely combines the R$^*$-tree with inverted files.

\noindent \emph{\textbf{IR$^2$-tree}}. A traditional spatio-textual index~\cite{defelipe2008icde} that tightly combines the R$^*$-tree with signatures encoding the textual information. As signatures, we use bitmaps with length equal to the number of distinct keywords per dataset.

\noindent \emph{\textbf{RSMI-IF}}. The index presented in Section~\ref{subsec:loosely}, which loosely combines RSMI with inverted files.

\noindent \emph{\textbf{RSMI-BM}}. The index presented in Section~\ref{subsec:tightly}, which tightly combines RSMI with bitmaps.

\noindent \emph{\textbf{RSMI-BM$^*$}}. The variant of RSMI-BM using the modified partitioning strategy in Section~\ref{subsec:tightly}.

\noindent \emph{\textbf{RSMI-BM-IR$^2$}}. The index presented in Section~\ref{subsec:hybrid}, which combines RSMI-BM with the non-learned IR$^2$-tree.

All methods were implemented in GNU C++. All indices are held in memory. The experiments were executed using gcc version 9.4.0 on a server with AMD Ryzen Threadripper 3960X 24-Core processor and 256 GB RAM running Ubuntu 20.04.1 LTS. 


\subsubsection{Index Parameters}
\label{subsubsec:params}
We conducted some preliminary tuning tests, and we have set the parameters of each index as described next.
The block size for RSMI-BM and RSMI-BM$^*$ was set to $|B|= 100$. Similarly, for the IR$^2$-Tree, we set the minimum and maximum node capacity to $m=50$ and $M=100$, respectively. For both R$^*$-Tree-IF and RSMI-IF, we set these values to $|B|= 1000$, $m=500$, and $M=1000$, to benefit from the inverted indices in the leaves. For RSMI-BM-IR$^2$, the block size was set to $|B|=1000$, while using $m=10$ and $M=20$ for the IR$^2$-trees constructed in the leaves in order to increase the tree height. For RSMI-IF, RSMI-BM, RSMI-BM$^*$ and RSMI-BM-IR$^2$ 
the maximum partition size was set to $S = 0.1 \cdot |\mathcal{D}|$ for each dataset.

The ML models used in the RSMI-based indices are multi-layer perceptrons (MLP) and their configuration is the same as in the original RSMI paper~\cite{qi2020effectively}.
Specifically, the hidden layer size is set to the sum of the number of dimensions and the number of output classes divided by two. For the hidden layer, we use the sigmoid activation function. The models are trained level by level, starting from the root, with the learning rate set to 0.01 and the number of epochs to 500.


\begin{table}[t]
    \centering
	\caption{Query parameters and their ranges.}
	\begin{tabular}{l|c}
	{\em \textbf{Parameter}} & {\em \textbf{Value}} \\
	\hline
	Window side length $\ell$ in BWQ queries & 0.01, 0.05, \textbf{0.1}, 0.15, 0.2  \\
    Count $k$ of results in B$k$Q queries & 5, \textbf{10}, 20, 50, 100  \\
    Number of query keywords $\mathcal{T}$ & 1, 2, 3, 4, 5  \\
	\end{tabular}
	\label{tab:params}
\end{table}



\subsubsection{Query Workload}
\label{subsubsec:queries}

The query parameters are listed in Table~\ref{tab:params}, with default values shown in bold. Each experiment is performed using a randomly selected workload of 1,000 objects as queries from each dataset, reporting the average response time. For BWQ, we construct the query windows as squares having as center the query location and specifying a varying side length $\ell$ as listed in Table~\ref{tab:params}. The window side length $\ell$ is expressed as a percentage of the whole dataset area. For tests involving a varying number of keywords $\mathcal{T}$, we isolate the spatio-textual objects containing at least $|\mathcal{T}|$ keywords in each case, and randomly select 1,000 objects as the respective query workload. To assess the quality of query results, we measure the average recall over the query workload. For B$k$Q queries, we also calculate the distance of the returned $k$NN from the query point and we measure its deviation (\%) from the distance of the exact $k$NN.


\subsection{Index Construction}
\label{subsec:index_construction}

\begin{figure}[!t]
\centering
\subfloat{\includegraphics[width=0.5\linewidth]{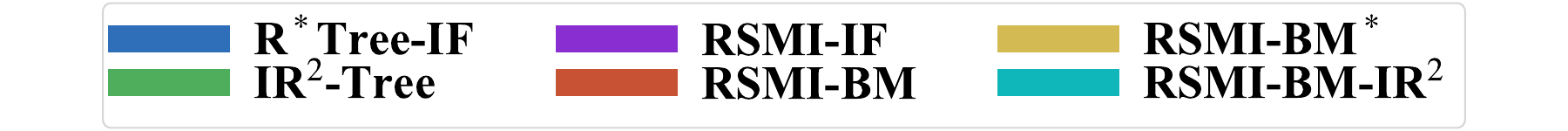}}
\\
\vspace{-15pt}
\setcounter{subfigure}{0}
\subfloat[Index build time.]{\includegraphics[trim=0.12cm 0.12cm 0.12cm 0.12cm, clip, width=0.35\textwidth]{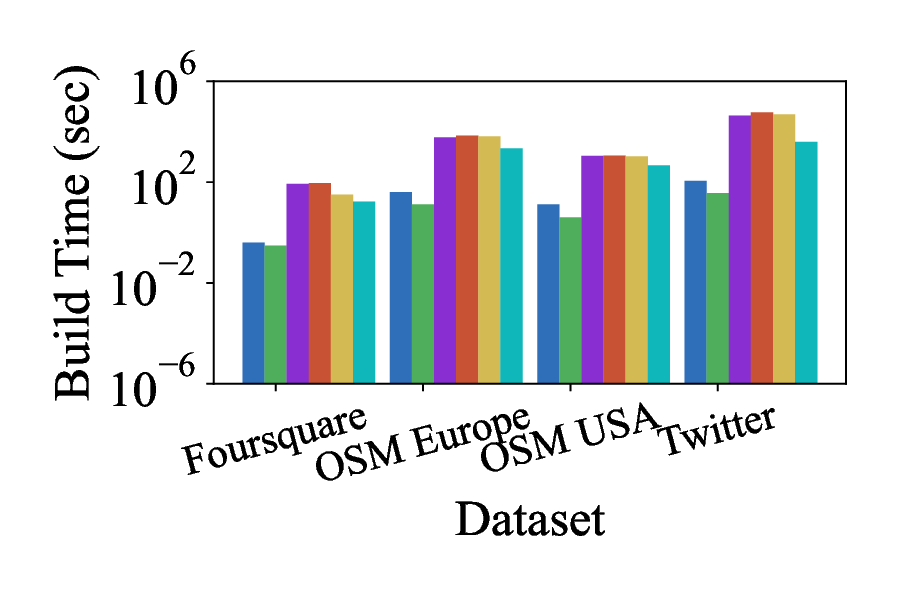}\label{subfig:index_build_time}}
\subfloat[Index size.]{\includegraphics[trim=0.12cm 0.12cm 0.12cm 0.12cm, clip, width=0.35\textwidth]{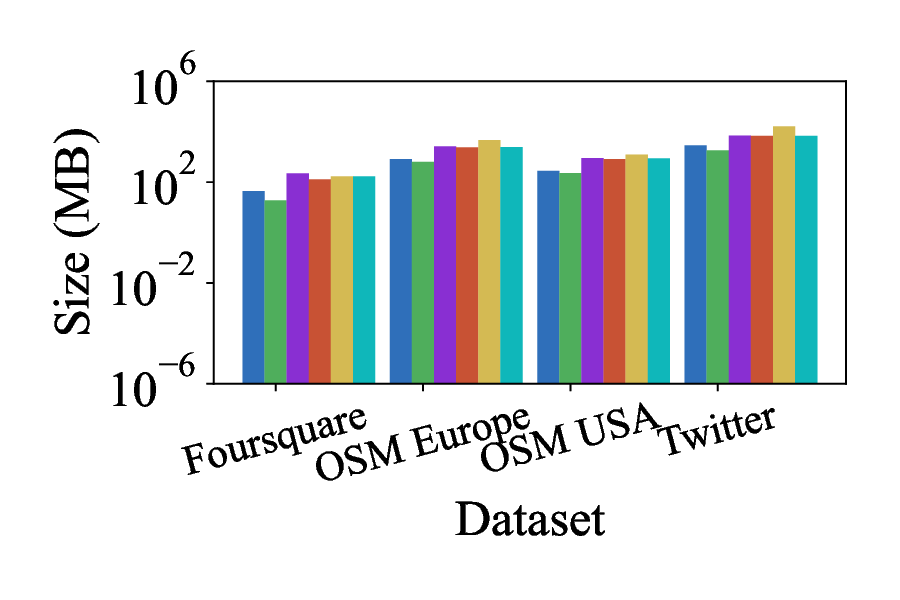}\label{subfig:index_size}}
\caption{Index build time and size.}
\label{fig:build_time_size}
\end{figure}

Figure~\ref{fig:build_time_size} depicts the build time and memory footprint required by each index per dataset. It is noticeable that traditional indices (i.e., IR$^2$-Tree and R*-Tree-IF) incur a significantly less construction cost, built almost an order of magnitude faster compared to the RSMI-based indices (Figure~\ref{subfig:index_build_time}). This is expected, since all proposed RSMI-based indices must undergo multiple expensive multi-epoch training tasks during construction; the more the models are trained, the more the construction cost.

Besides, traditional indices consume less memory (Figure~\ref{subfig:index_size}).  The increased memory footprint of the RSMI-based indices should be attributed to the ML models that must also be kept in memory and the large number of bitmaps stored per leaf node and block along with the raw data (point coordinates and keywords). Furthermore, the RSMI-based indices by default must keep for each point the mapped coordinates to the rank space and the space-filling curve value, thus requiring extra memory.



\subsection{BWQ Performance Results}
\label{subsec:window_queries}

\begin{figure*}[!h]
\centering
\subfloat{\includegraphics[width=\linewidth]{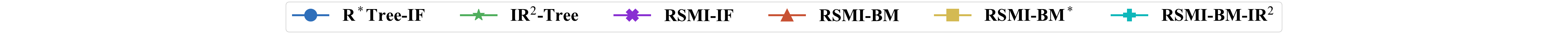}}
\\
\vspace{-10pt}
\setcounter{subfigure}{0}
\subfloat[Foursquare Dataset]{\includegraphics[trim=0.12cm 0.12cm 0.12cm 0.12cm, clip, width=0.25\textwidth]{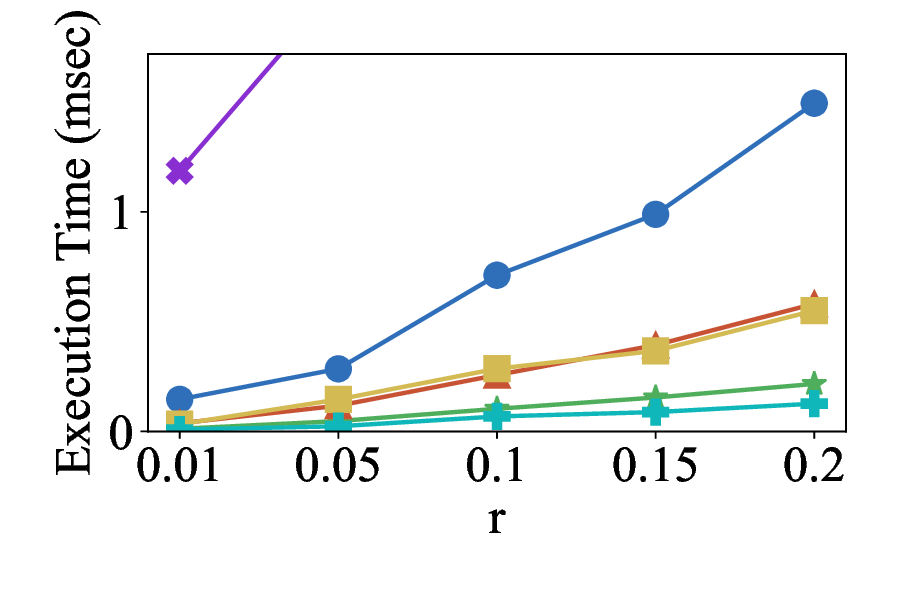}\label{subfig:var_r1}}
\subfloat[OSM USA Dataset]{\includegraphics[trim=0.12cm 0.12cm 0.12cm 0.12cm, clip, width=0.25\textwidth]{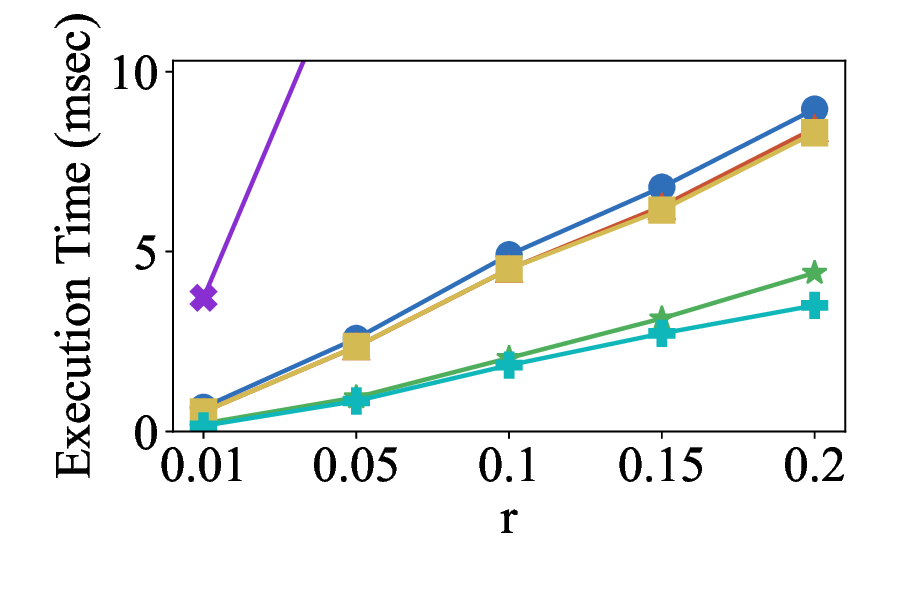}\label{subfig:var_r4}}
\subfloat[OSM Europe Dataset]{\includegraphics[trim=0.12cm 0.12cm 0.12cm 0.12cm, clip, width=0.25\textwidth]{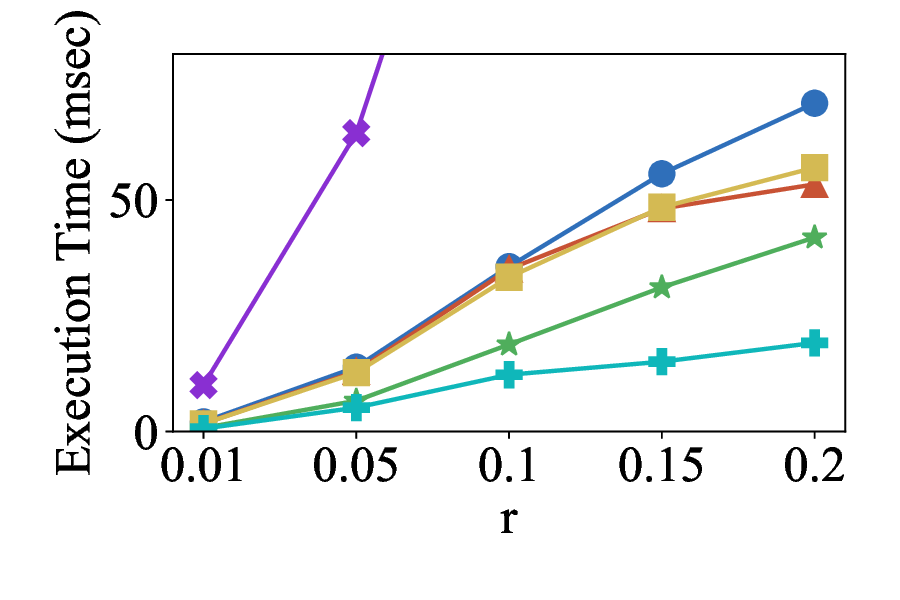}\label{subfig:var_r2}}
\subfloat[Twitter Dataset]{\includegraphics[trim=0.12cm 0.12cm 0.12cm 0.12cm, clip, width=0.25\textwidth]{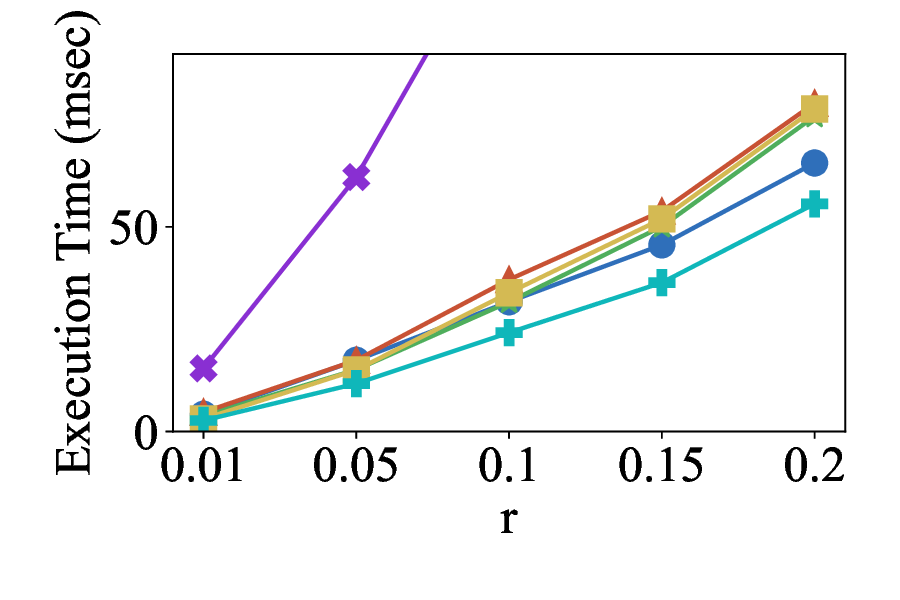}\label{subfig:var_r3}}
\caption{BWQ response time with varying window size.}
\label{fig:varying_r}
\end{figure*}

\begin{figure*}[!t]
\centering
\subfloat{\includegraphics[width=\linewidth]{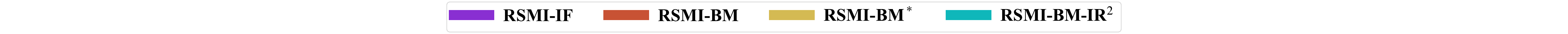}}
\\
\vspace{-10pt}
\setcounter{subfigure}{0}
\subfloat[Foursquare Dataset]{\includegraphics[trim=0.12cm 0.12cm 0.12cm 0.12cm, clip, width=0.25\textwidth]{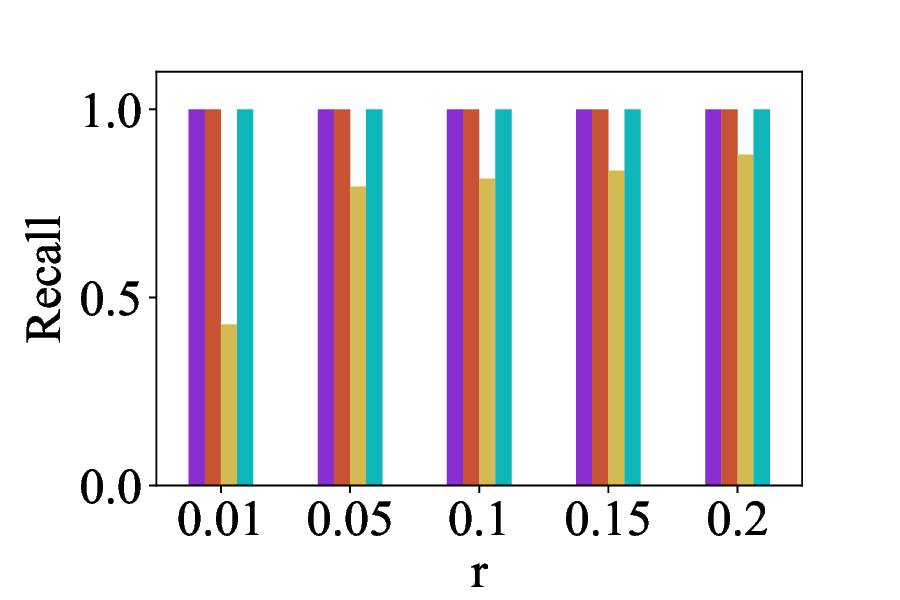}\label{subfig:var_r1_r}}
\subfloat[OSM USA Dataset]{\includegraphics[trim=0.12cm 0.12cm 0.12cm 0.12cm, clip, width=0.25\textwidth]{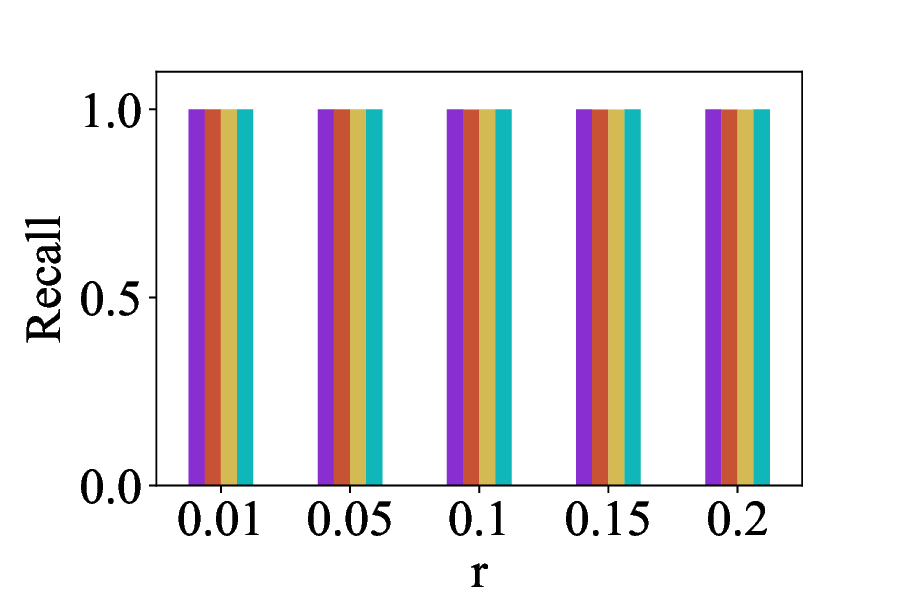}\label{subfig:var_r4_r}}
\subfloat[OSM Europe Dataset]{\includegraphics[trim=0.12cm 0.12cm 0.12cm 0.12cm, clip, width=0.25\textwidth]{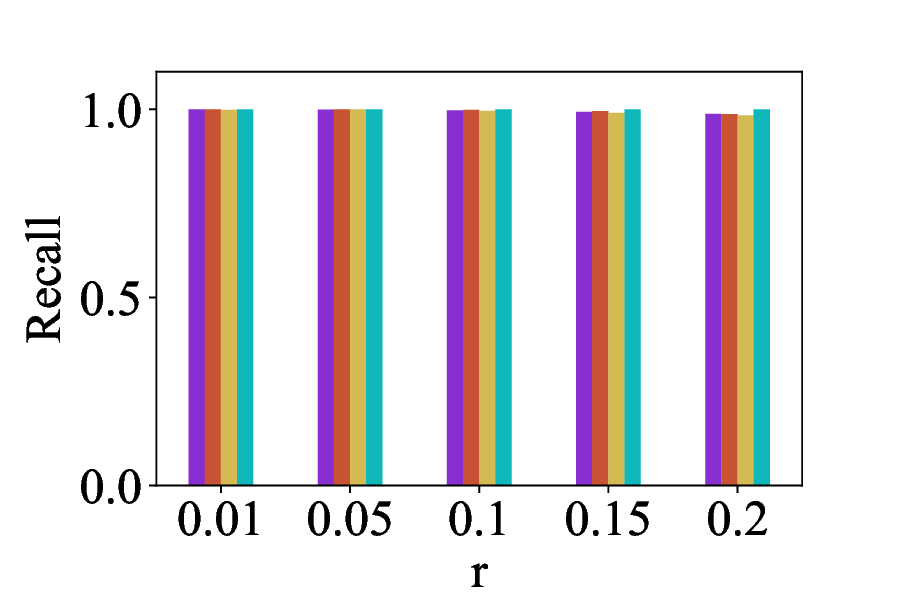}\label{subfig:var_r2_r}}
\subfloat[Twitter Dataset]{\includegraphics[trim=0.12cm 0.12cm 0.12cm 0.12cm, clip, width=0.25\textwidth]{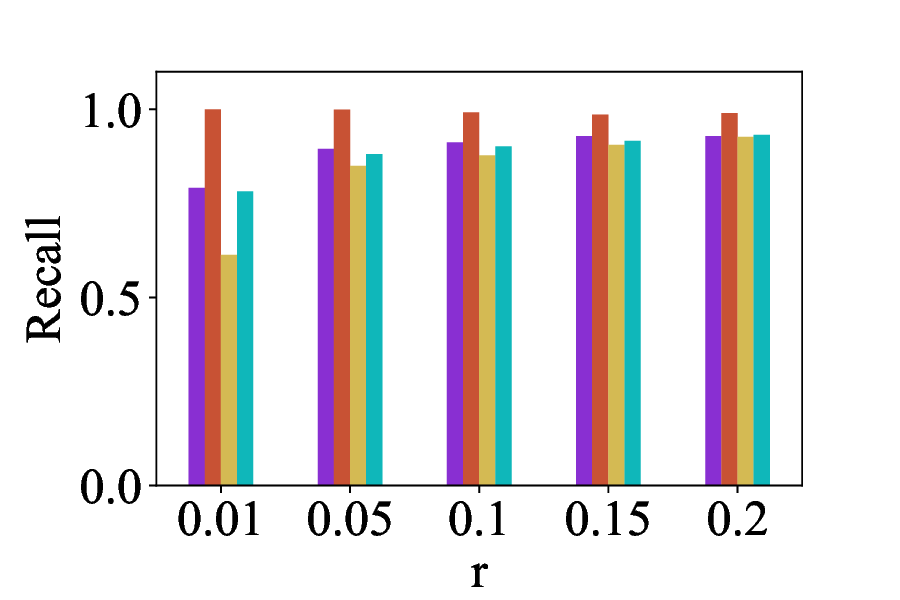}\label{subfig:var_r3_r}}
\caption{Recall of BWQ results for varying window sizes.}
\label{fig:varying_r_r}
\end{figure*}

\paragraph{Varying Window Size} Figure~\ref{fig:varying_r} illustrates performance of BWQ for various window sizes. Unsurprisingly, RSMI-IF is significantly slow, an order of magnitude worse than its competitors, as it does not take into account the query keywords while traversing the RSMI. It also has to perform multiple additional searches in inverted files of the leaves followed by expensive set intersection operations.
The fastest scheme in all cases is RSMI-BM-IR$^2$, which manages to cope well, taking advantage of RSMI's quick node traversal. IR$^2$-Tree comes in a close second in most cases, while all other indices are less efficient. Interestingly, though, the R*-Tree-IF manages to outperform IR$^2$-Tree for larger window sizes against the Twitter dataset, despite being the slowest in all other cases. This is possibly due to the larger node capacities along with efficient pruning on the spatial domain, requiring only few interactions with the inverted files at the leaf level.

The corresponding recall results are shown in Figure~\ref{fig:varying_r_r}. Noticeably, recall values for the RSMI-based methods on OSM data are all close to 1. Performance of RSMI-BM$^*$ against the Foursquare data is poor, especially for smaller window sizes, possibly because the ML models are struggling to learn the calculated partitions. Interestingly, for the Twitter dataset, RSMI-BM achieves the highest recall (close to 1) for all window sizes, possibly due to a partitioning that was easier to learn. In general, we noticed that RSMI-based indices with multiple levels tend to have many false negatives, since inner node traversal does not take into account the model errors. 

\begin{figure*}[!h]
\centering
\subfloat{\includegraphics[width=\linewidth]{Figures/legend.pdf}}
\\
\vspace{-10pt}
\setcounter{subfigure}{0}
\subfloat[Foursquare Dataset]{\includegraphics[trim=0.12cm 0.12cm 0.12cm 0.12cm, clip, width=0.25\textwidth]{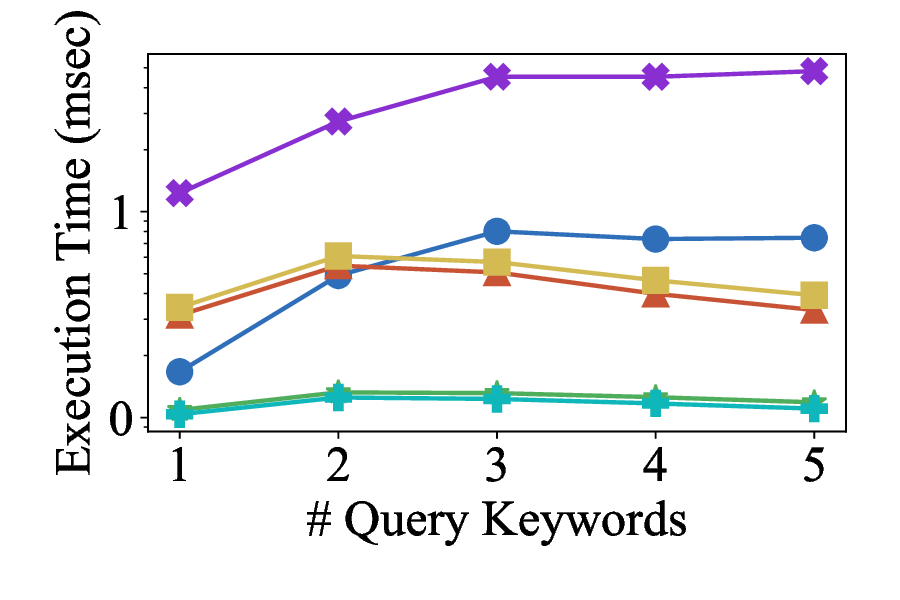}\label{subfig:var_w1}}
\subfloat[OSM USA Dataset]{\includegraphics[trim=0.12cm 0.12cm 0.12cm 0.12cm, clip, width=0.25\textwidth]{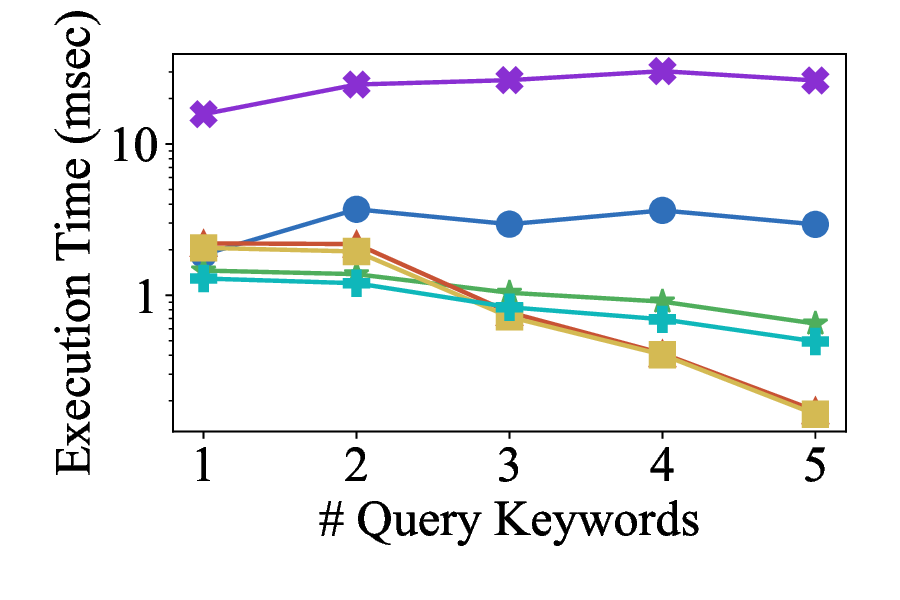}\label{subfig:var_w4}}
\subfloat[OSM Europe Dataset]{\includegraphics[trim=0.12cm 0.12cm 0.12cm 0.12cm, clip, width=0.25\textwidth]{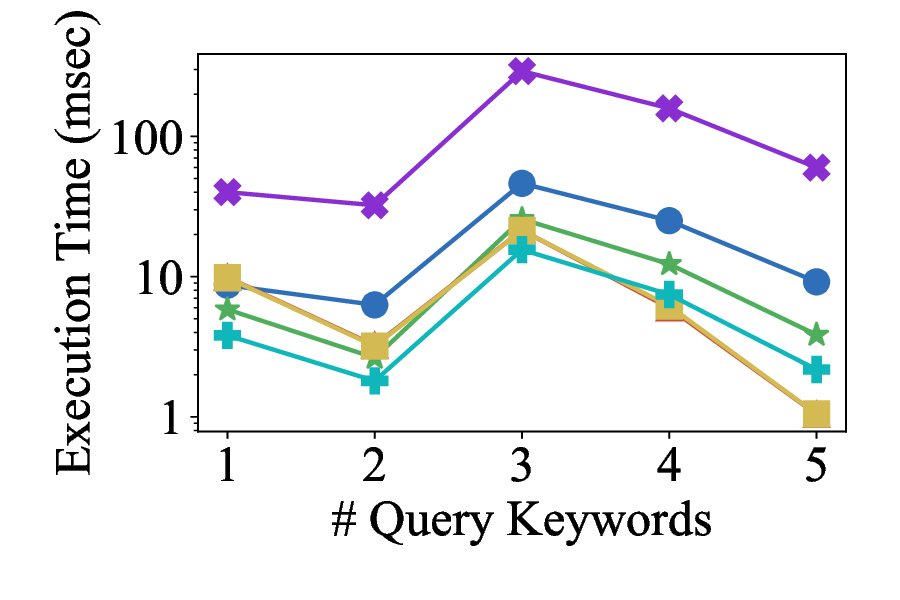}\label{subfig:var_w2}}
\subfloat[Twitter Dataset]{\includegraphics[trim=0.12cm 0.12cm 0.12cm 0.12cm, clip, width=0.25\textwidth]{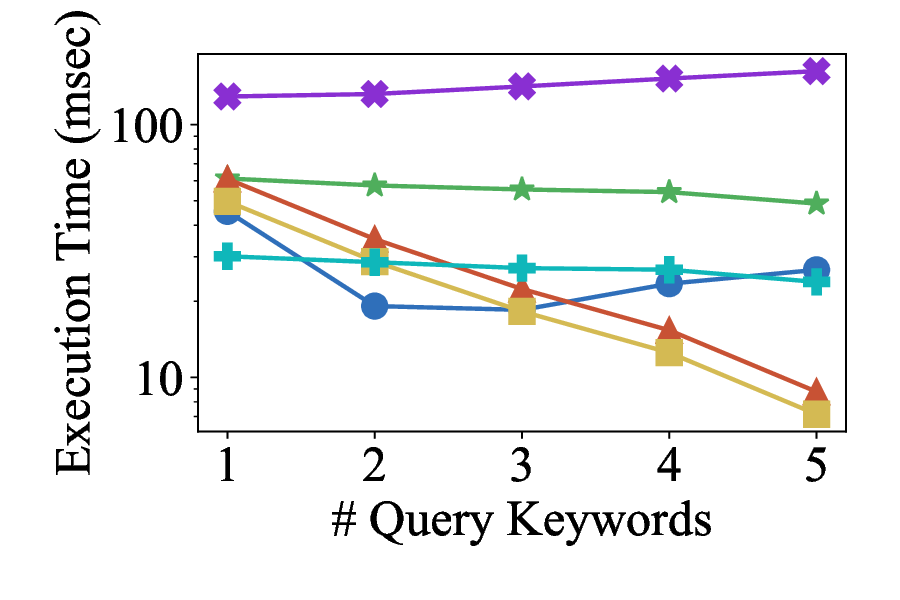}\label{subfig:var_w3}}
\caption{BWQ response time with varying number of query keywords $\mathcal{T}$.}
\label{fig:varying_w}
\end{figure*}

\begin{figure*}[!h]
\centering
\subfloat{\includegraphics[width=\linewidth]{Figures/legend4.pdf}}
\\
\vspace{-10pt}
\setcounter{subfigure}{0}
\subfloat[Foursquare Dataset]{\includegraphics[trim=0.12cm 0.12cm 0.12cm 0.12cm, clip, width=0.25\textwidth]{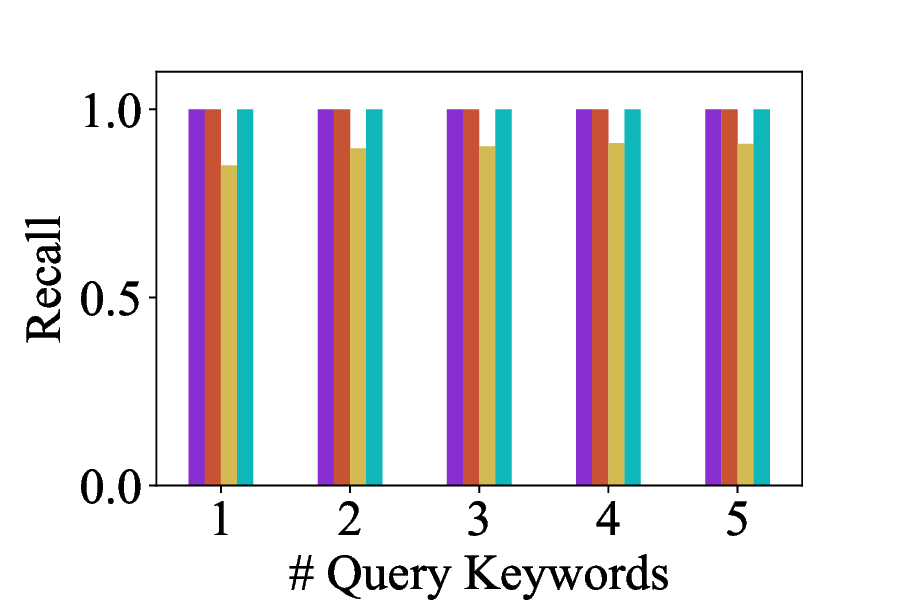}\label{subfig:var_w1_r}}
\subfloat[OSM USA Dataset]{\includegraphics[trim=0.12cm 0.12cm 0.12cm 0.12cm, clip, width=0.25\textwidth]{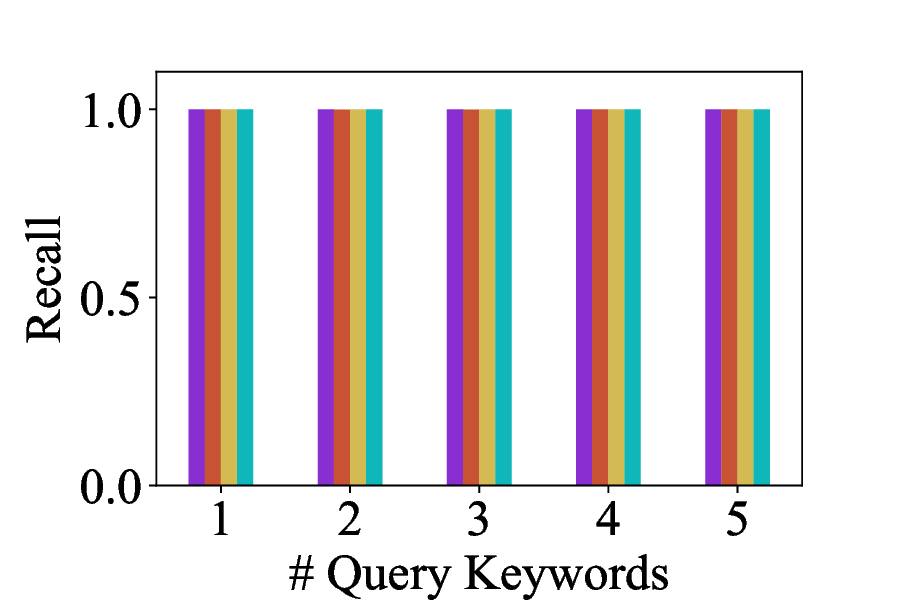}\label{subfig:var_w4_r}}
\subfloat[OSM Europe Dataset]{\includegraphics[trim=0.12cm 0.12cm 0.12cm 0.12cm, clip, width=0.25\textwidth]{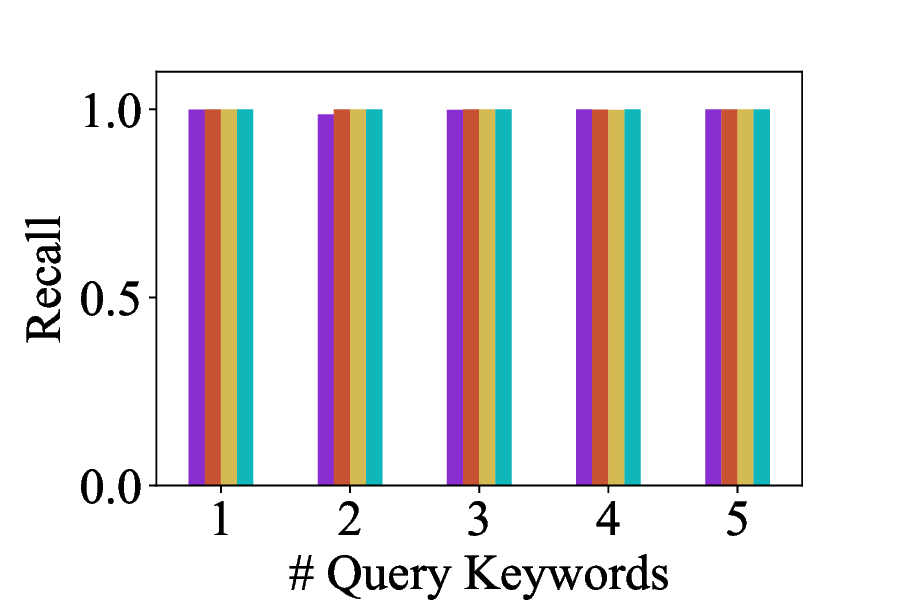}\label{subfig:var_w2_r}}
\subfloat[Twitter Dataset]{\includegraphics[trim=0.12cm 0.12cm 0.12cm 0.12cm, clip, width=0.25\textwidth]{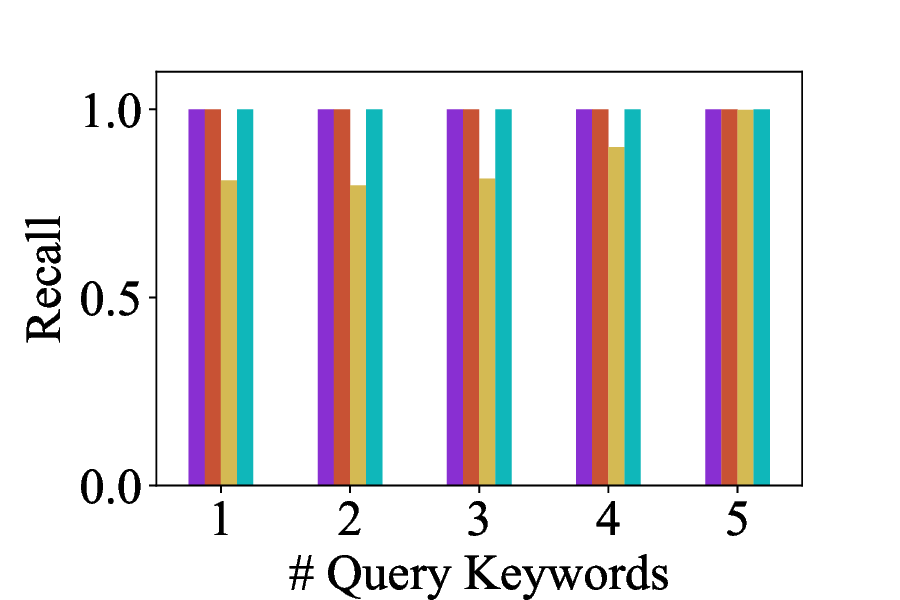}\label{subfig:var_w3_r}}
\caption{Recall of BWQ results for varying number of query keywords $\mathcal{T}$.}
\label{fig:varying_w_r}
\end{figure*}

\paragraph{Varying Number of Query Keywords} Figure~\ref{fig:varying_w} plots BWQ response times for a varying number $|\mathcal{T}|$ of keywords per query. As previously noted, RSMI-IF is the slowest scheme. Since objects with many keywords are rather rare (see Table~\ref{tab:datasets}), increasing the number of keywords decreases the selectivity in all methods and performance is either improved, or remains stable. In general, RSMI-BM-IR$^2$ is again the fastest; however, it is outperformed by RSMI-BM and RSMI-BM$^*$ for $|\mathcal{T}| > 3$, except for the small Foursquare dataset.

Regarding accuracy, note that recall is almost equal to 1 in most cases (Figure~\ref{fig:varying_w_r}). The only exception is RSMI-BM$^*$, which offers less accurate results in some settings over the Foursquare and Twitter datasets, possibly because it is harder to learn the derived partitions. Nevertheless, for queries with $|\mathcal{T}| > 4$ keywords over the Twitter dataset, the RSMI-BM$^*$ approach is both the fastest and achieves a very high recall close to 1.

\subsection{B$k$Q Performance Results}
\label{subsec:kNN_queries}

\begin{figure*}[!h]
\centering
\subfloat{\includegraphics[width=\linewidth]{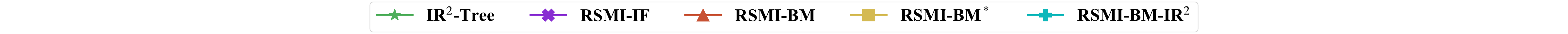}}
\\
\vspace{-10pt}
\setcounter{subfigure}{0}
\subfloat[Foursquare Dataset]{\includegraphics[trim=0.12cm 0.12cm 0.12cm 0.12cm, clip, width=0.25\textwidth]{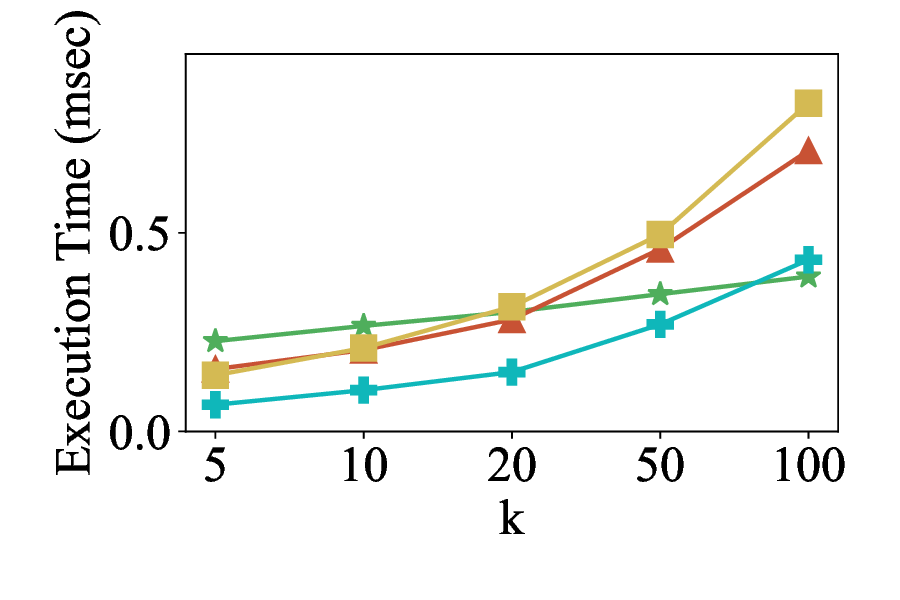}\label{subfig:var_k1}}
\subfloat[OSM USA Dataset]{\includegraphics[trim=0.12cm 0.12cm 0.12cm 0.12cm, clip, width=0.25\textwidth]{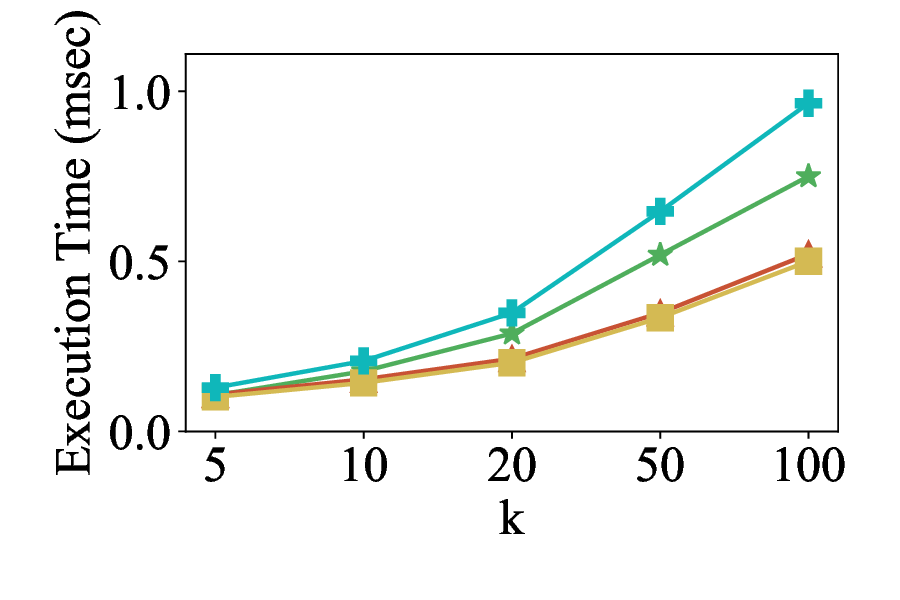}\label{subfig:var_k4}}
\subfloat[OSM Europe Dataset]{\includegraphics[trim=0.12cm 0.12cm 0.12cm 0.12cm, clip, width=0.25\textwidth]{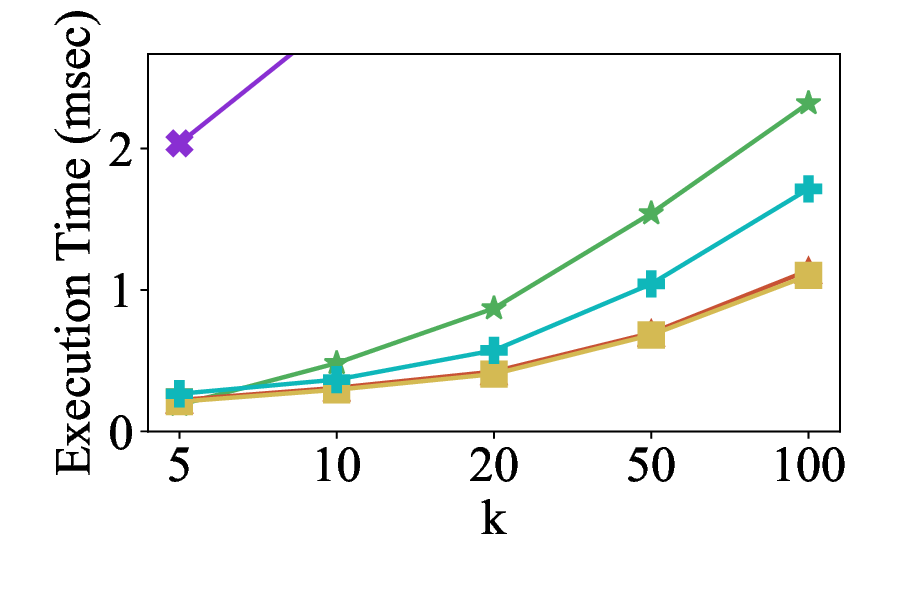}\label{subfig:var_k2}}
\subfloat[Twitter Dataset]{\includegraphics[trim=0.12cm 0.12cm 0.12cm 0.12cm, clip, width=0.25\textwidth]{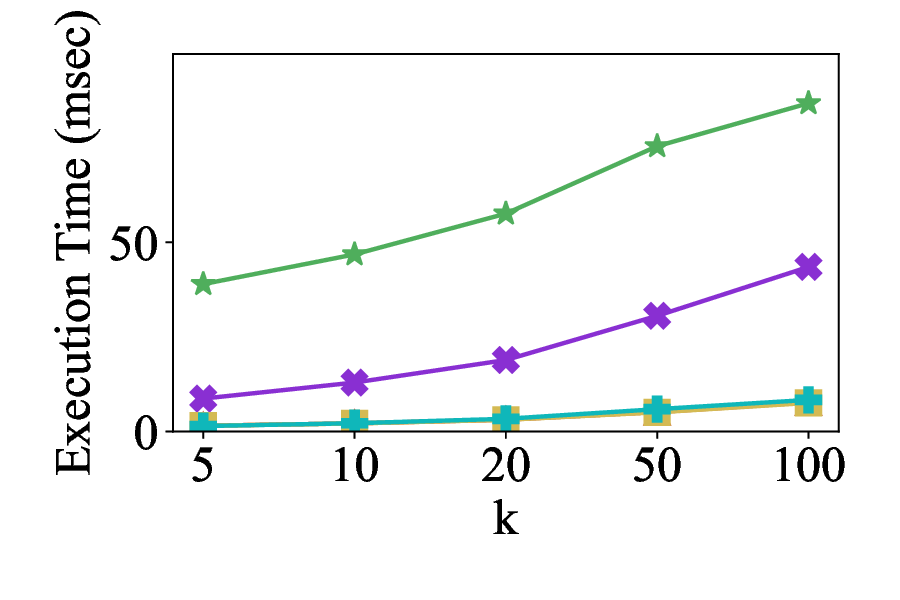}\label{subfig:var_k3}}
\caption{B$k$Q response time for varying $k$.}
\label{fig:varying_k}
\end{figure*}

\begin{figure*}[!h]
\centering
\subfloat{\includegraphics[width=\linewidth]{Figures/legend4.pdf}}
\\
\vspace{-10pt}
\setcounter{subfigure}{0}
\subfloat[Foursquare Dataset]{\includegraphics[trim=0.12cm 0.12cm 0.12cm 0.12cm, clip, width=0.25\textwidth]{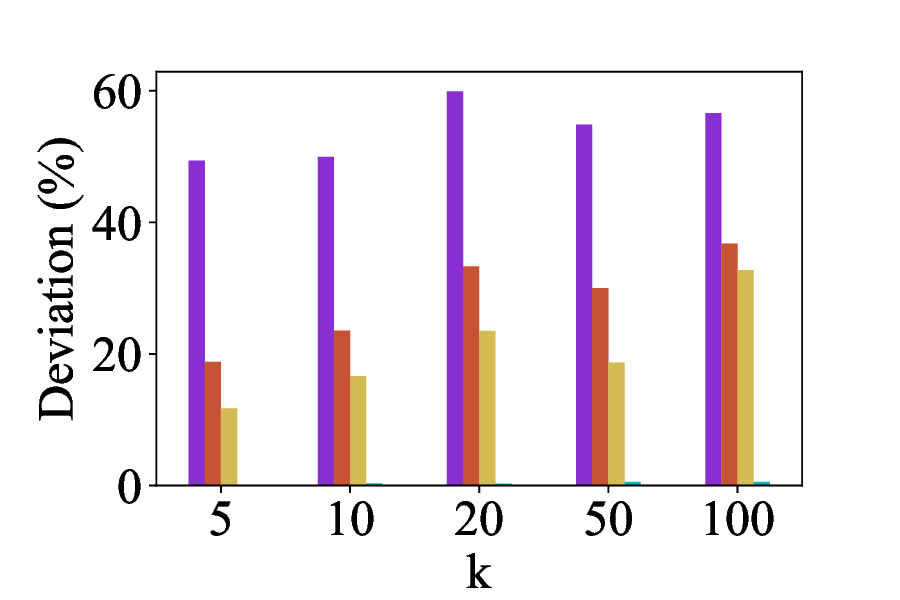}\label{subfig:var_k1_d}}
\subfloat[OSM USA Dataset]{\includegraphics[trim=0.12cm 0.12cm 0.12cm 0.12cm, clip, width=0.25\textwidth]{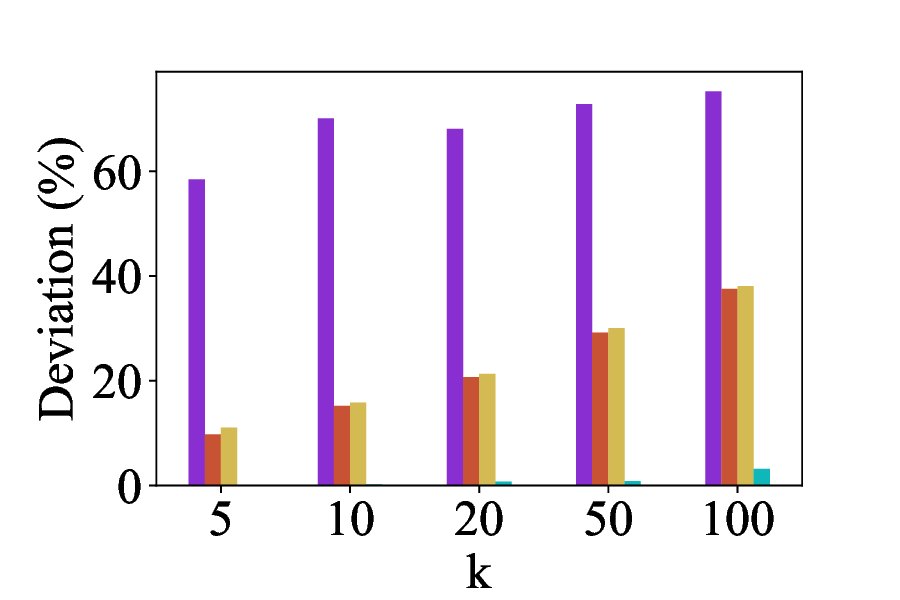}\label{subfig:var_k4_d}}
\subfloat[OSM Europe Dataset]{\includegraphics[trim=0.12cm 0.12cm 0.12cm 0.12cm, clip, width=0.25\textwidth]{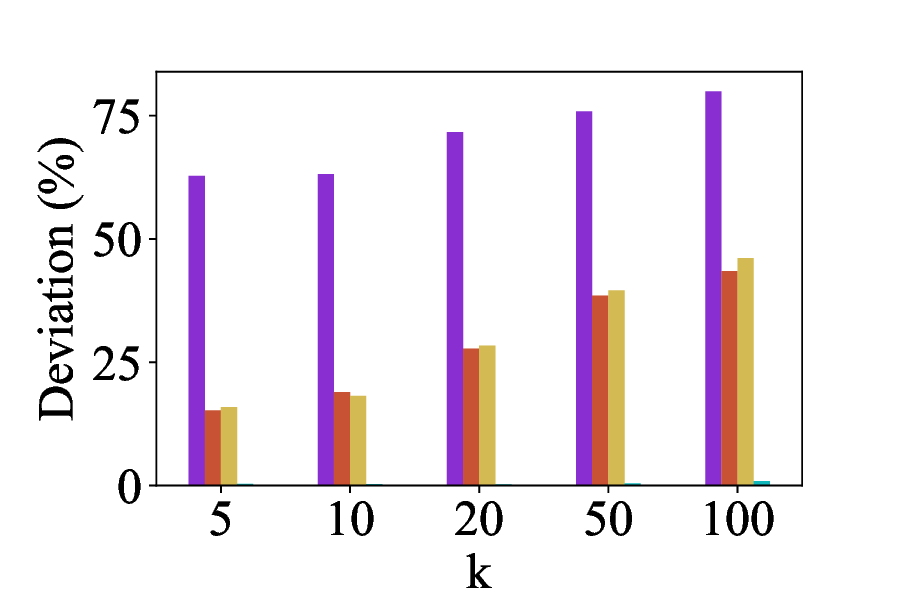}\label{subfig:var_k2_d}}
\subfloat[Twitter Dataset]{\includegraphics[trim=0.12cm 0.12cm 0.12cm 0.12cm, clip, width=0.25\textwidth]{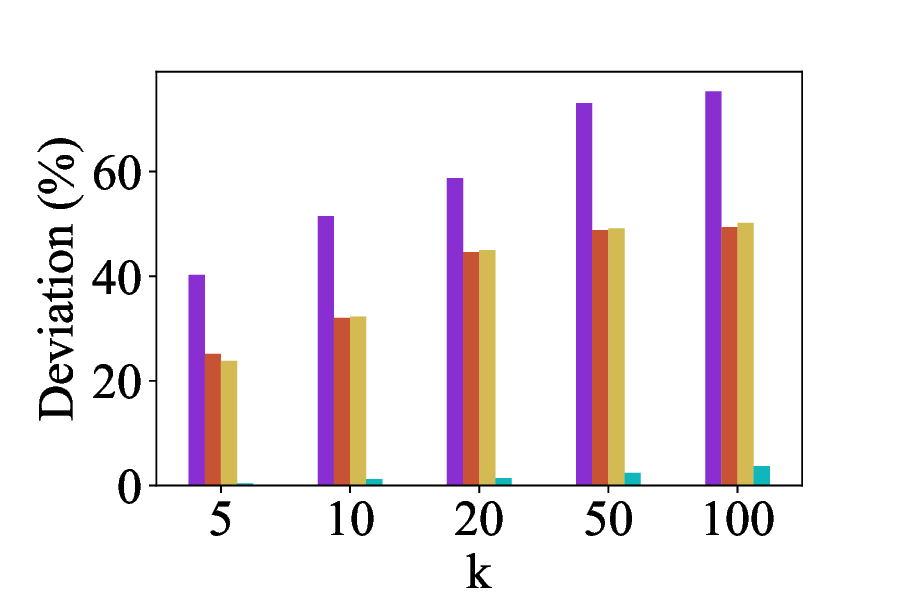}\label{subfig:var_k3_d}}
\caption{Distance deviation of $k$-th NN in B$k$Q results for varying $k$.}
\label{fig:varying_k_d}
\end{figure*}

\begin{figure*}[!h]
\centering
\subfloat{\includegraphics[width=\linewidth]{Figures/legend4.pdf}}
\\
\vspace{-10pt}
\setcounter{subfigure}{0}
\subfloat[Foursquare Dataset]{\includegraphics[trim=0.12cm 0.12cm 0.12cm 0.12cm, clip, width=0.25\textwidth]{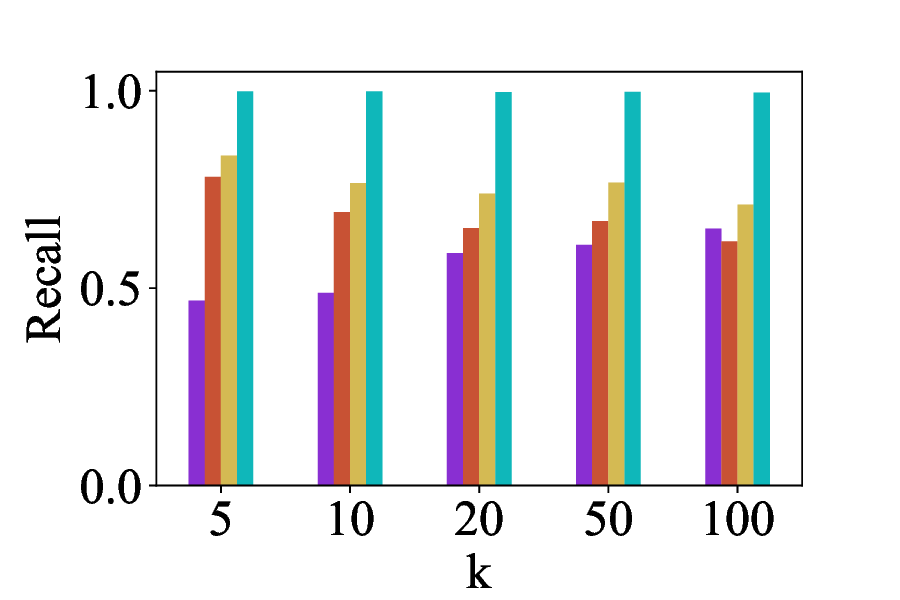}\label{subfig:var_k1_r}}
\subfloat[OSM USA Dataset]{\includegraphics[trim=0.12cm 0.12cm 0.12cm 0.12cm, clip, width=0.25\textwidth]{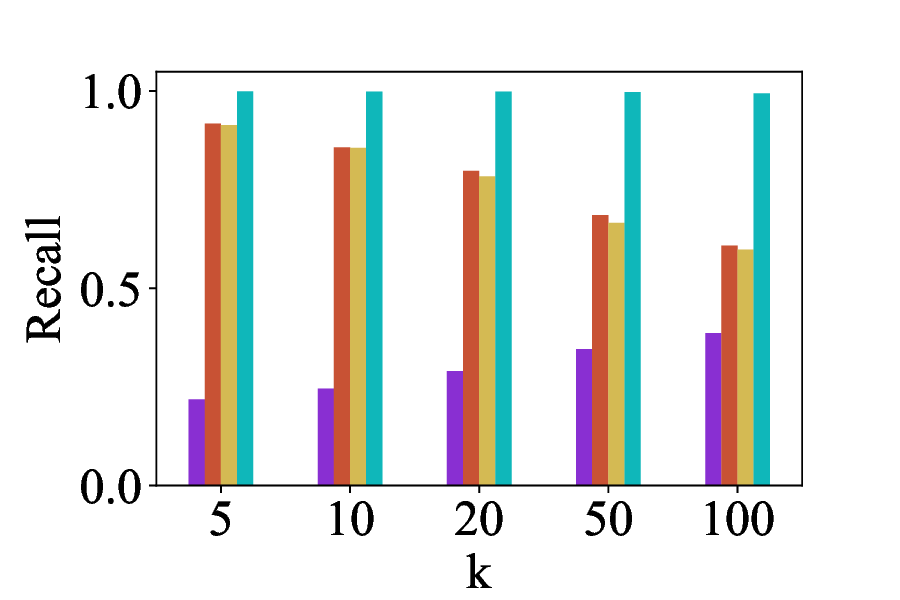}\label{subfig:var_k4_r}}
\subfloat[OSM Europe Dataset]{\includegraphics[trim=0.12cm 0.12cm 0.12cm 0.12cm, clip, width=0.25\textwidth]{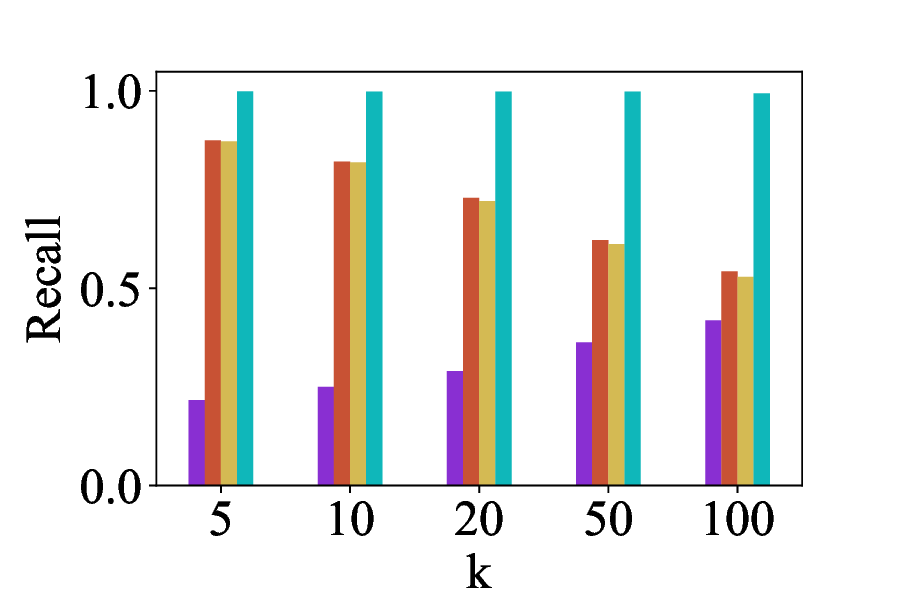}\label{subfig:var_k2_r}}
\subfloat[Twitter Dataset]{\includegraphics[trim=0.12cm 0.12cm 0.12cm 0.12cm, clip, width=0.25\textwidth]{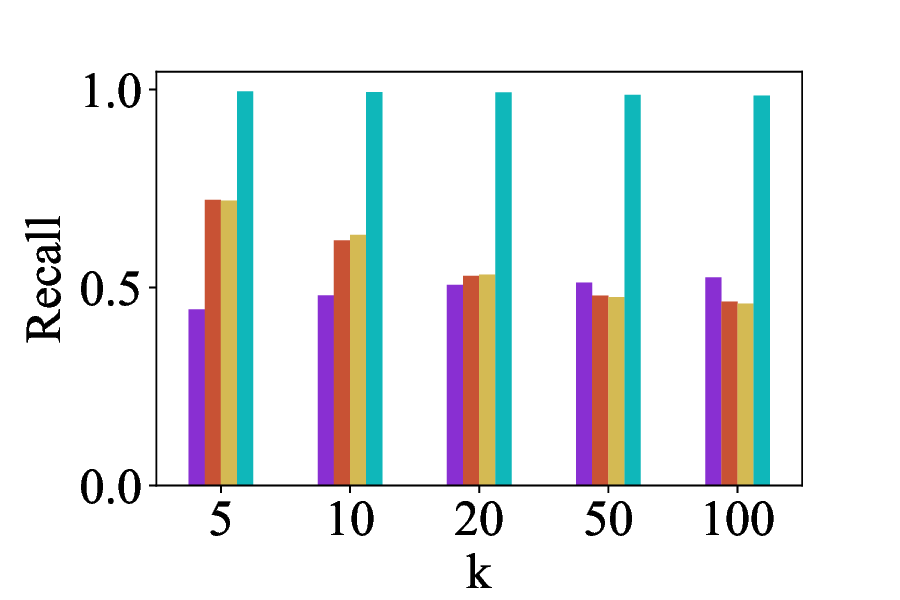}\label{subfig:var_k3_r}}
\caption{Recall of B$k$Q results for varying $k$.}
\label{fig:varying_k_r}
\end{figure*}

\paragraph{Varying $k$} Figure~\ref{fig:varying_k} illustrates B$k$Q performance for a varying number $k$ of returned results. Note that R$^*$-Tree-IF does not support such queries, hence it is not included in this experiment. In contrast to BWQ performance results, the RSMI-BM-IR$^2$ approach is now the fastest only over the Foursquare and Twitter datasets. For the rather small Foursquare dataset, the IR$^2$-Tree scales better for larger $k$, overtaking RSMI-BM-IR$^2$. However, this is not the case over the OSM datasets, where RSMI-BM and RSMI-BM$^*$ are almost on par (RSMI-BM$^*$ is slightly faster), and generally faster than their competitors. While RSMI-BM-IR$^2$ is faster than IR$^2$-Tree over the OSM Europe data, it struggles over OSM USA data. The most interesting observation, though, concerns the larger Twitter dataset, where IR$^2$-Tree is significantly slower than the RSMI-based indices. In this case, these latter learned schemes manage to detect the requested $k$ nearest neighbors without issuing many window queries.

Figure~\ref{fig:varying_k_d} depicts the distance deviation between exact and RSMI-based $k$-th nearest neighbors from the query point. Intuitively, this indicates the quality of the RSMI-based results by considering the distance difference from the farthest result. The loosely coupled RSMI-IF has poor accuracy, with distance deviations reaching 75\% in some cases. RSMI-BM$^*$ and RSMI-BM manage to cope reasonably well, with a deviation of 20-40\% in most cases. The best performance is achieved by RSMI-BM-IR$^2$, which yields an almost zero deviation in most cases. This observation is also confirmed by the corresponding recall values (Figure~\ref{fig:varying_k_r}), with RSMI-BM-IR$^2$ achieving a recall close to 1 in all cases. Now, the superiority of RSMI-BM-IR$^2$ is more evident; by taking advantage of the fast RSMI-based inner node traversal and the accurate query results of IR$^2$-Tree in the leaves, it manages to achieve very low response times in all cases, while maintaining a recall approximately equal to 1.

\begin{figure*}[!h]
\centering
\subfloat{\includegraphics[width=\linewidth]{Figures/legend1.pdf}}
\\
\vspace{-10pt}
\setcounter{subfigure}{0}
\subfloat[Foursquare Dataset]{\includegraphics[trim=0.12cm 0.12cm 0.12cm 0.12cm, clip, width=0.25\textwidth]{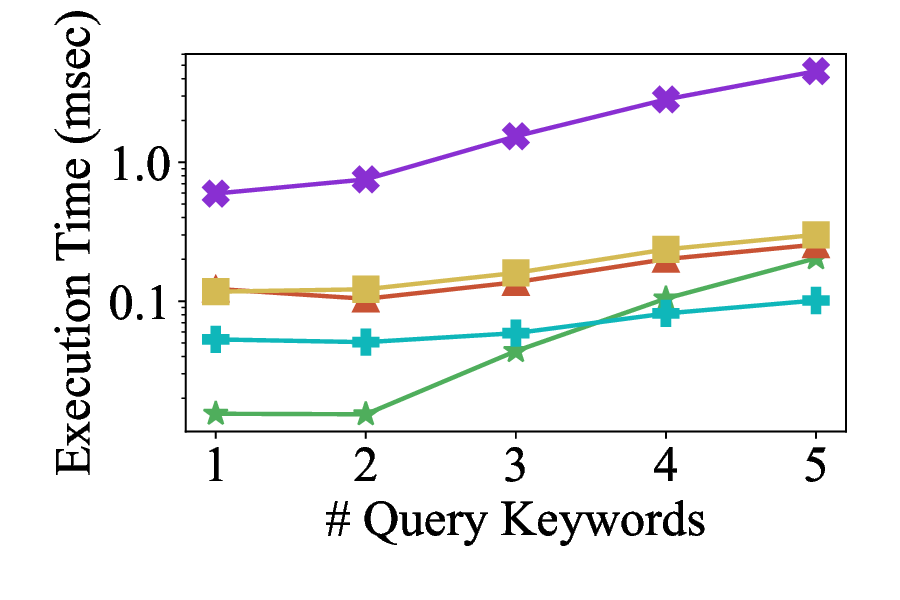}\label{subfig:var_w11}}
\subfloat[OSM USA Dataset]{\includegraphics[trim=0.12cm 0.12cm 0.12cm 0.12cm, clip, width=0.25\textwidth]{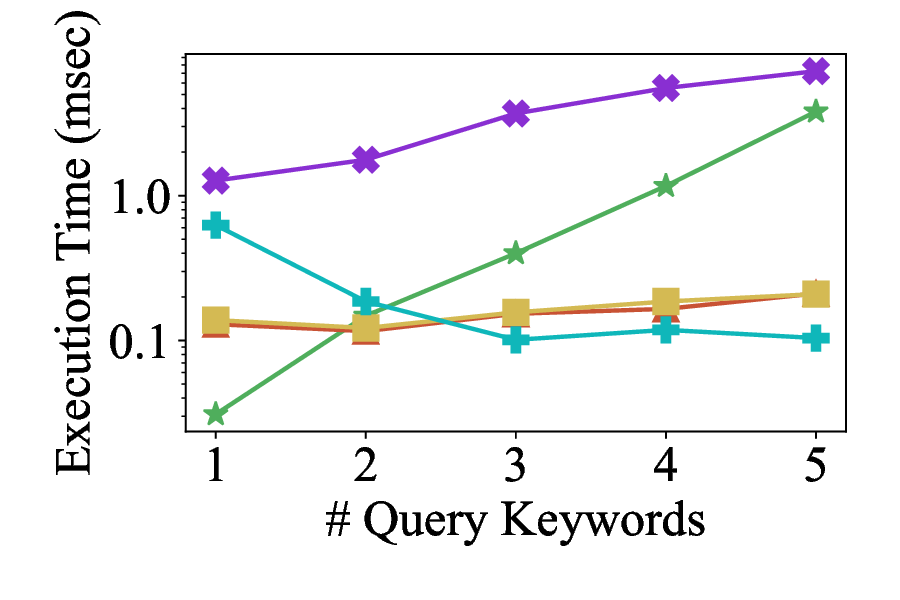}\label{subfig:var_w41}}
\subfloat[OSM Europe Dataset]{\includegraphics[trim=0.12cm 0.12cm 0.12cm 0.12cm, clip, width=0.25\textwidth]{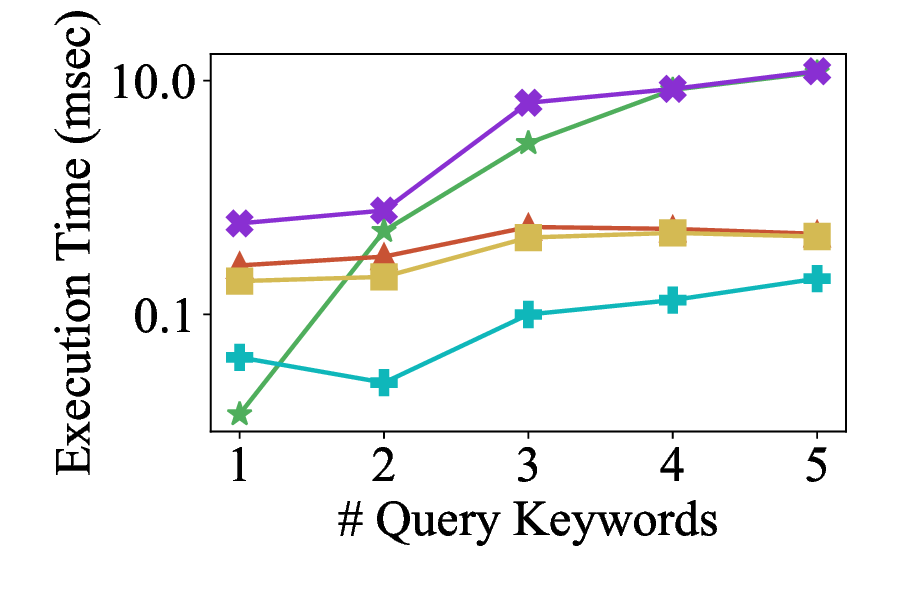}\label{subfig:var_w19}}
\subfloat[Twitter Dataset]{\includegraphics[trim=0.12cm 0.12cm 0.12cm 0.12cm, clip, width=0.25\textwidth]{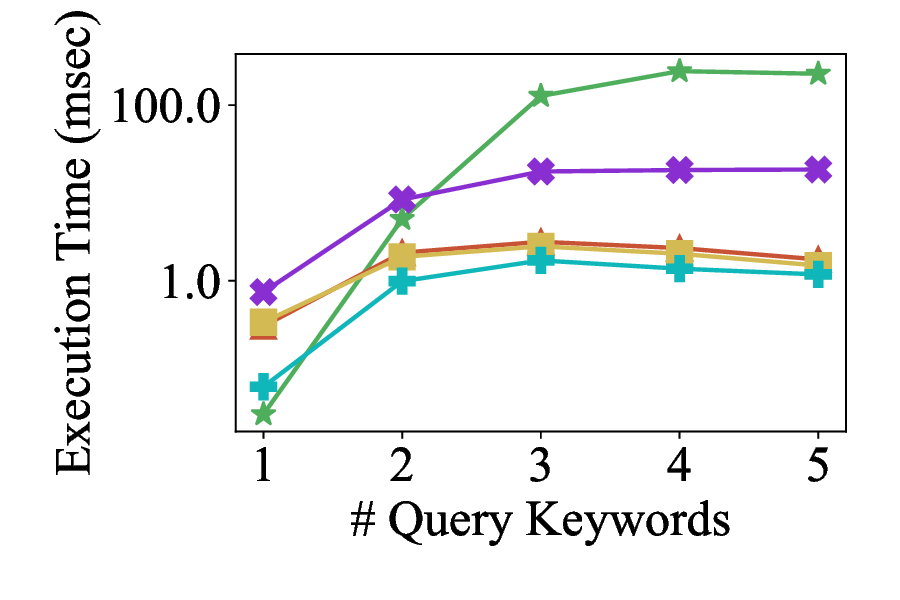}\label{subfig:var_w31}}
\caption{B$k$Q response time with varying number of query keywords $\mathcal{T}$.}
\label{fig:varying_w1}
\end{figure*}

\paragraph{Varying Number of Query Keywords} Figure~\ref{fig:varying_w1} illustrates B$k$Q performance for a varying number $|\mathcal{T}|$ of keywords per query. Increasing the number of keywords does not improve the response time, since queries with more keywords are rather rare and more searches are needed to fetch all qualifying $k$ neighbors. RSMI-BM-IR$^2$ is again the fastest in most cases. Interestingly, IR$^2$-Tree does not scale well; it is the fastest approach for $|\mathcal{T}|=1$ (or |$\mathcal{T}|=2$ over the Foursquare dataset), but it gets significantly slower with more query keywords. RSMI-based indices in general seem to cope well for any number of query keywords.

\begin{figure*}[!h]
\centering
\subfloat{\includegraphics[width=\linewidth]{Figures/legend4.pdf}}
\\
\vspace{-10pt}
\setcounter{subfigure}{0}
\subfloat[Foursquare Dataset]{\includegraphics[trim=0.12cm 0.12cm 0.12cm 0.12cm, clip, width=0.25\textwidth]{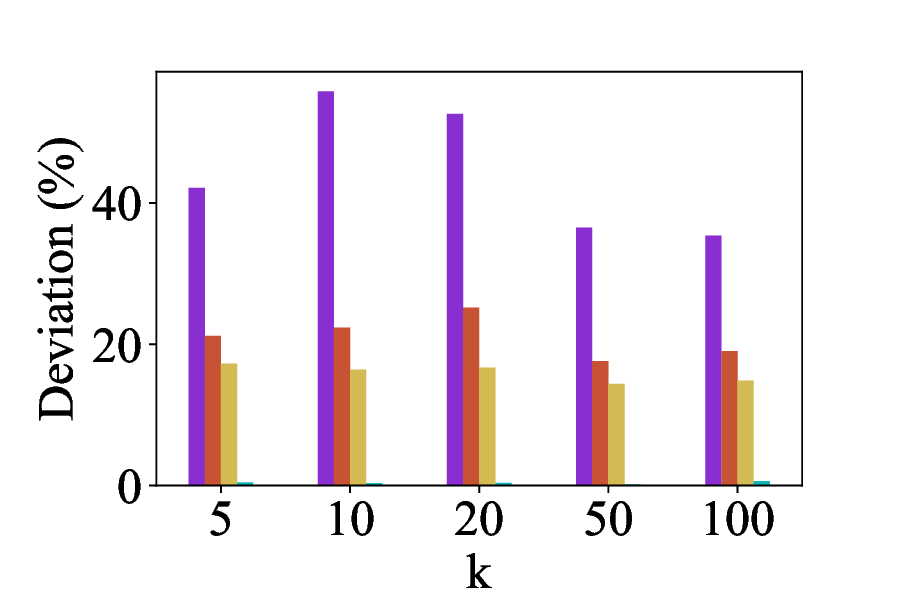}\label{subfig:var_w11_d}}
\subfloat[OSM USA Dataset]{\includegraphics[trim=0.12cm 0.12cm 0.12cm 0.12cm, clip, width=0.25\textwidth]{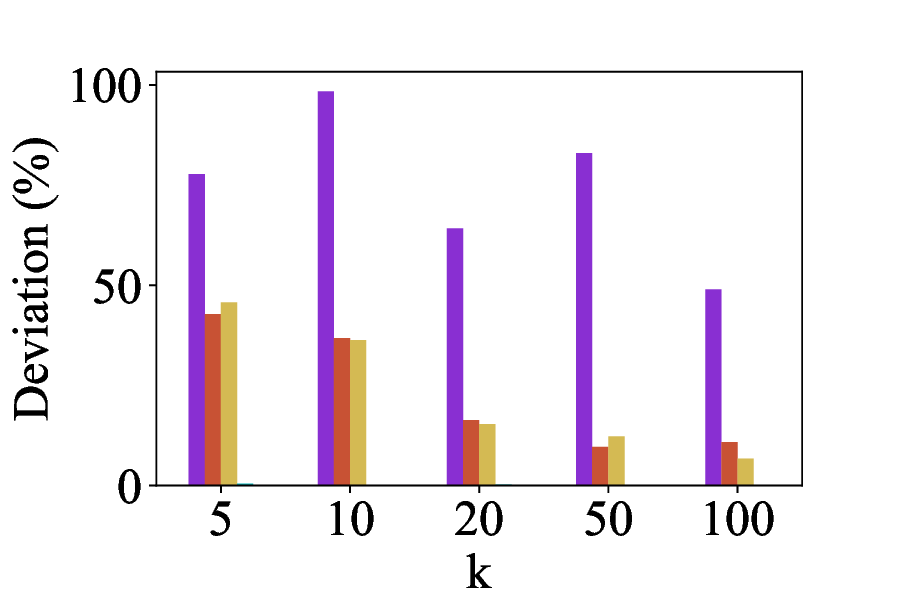}\label{subfig:var_w41_d}}
\subfloat[OSM Europe Dataset]{\includegraphics[trim=0.12cm 0.12cm 0.12cm 0.12cm, clip, width=0.25\textwidth]{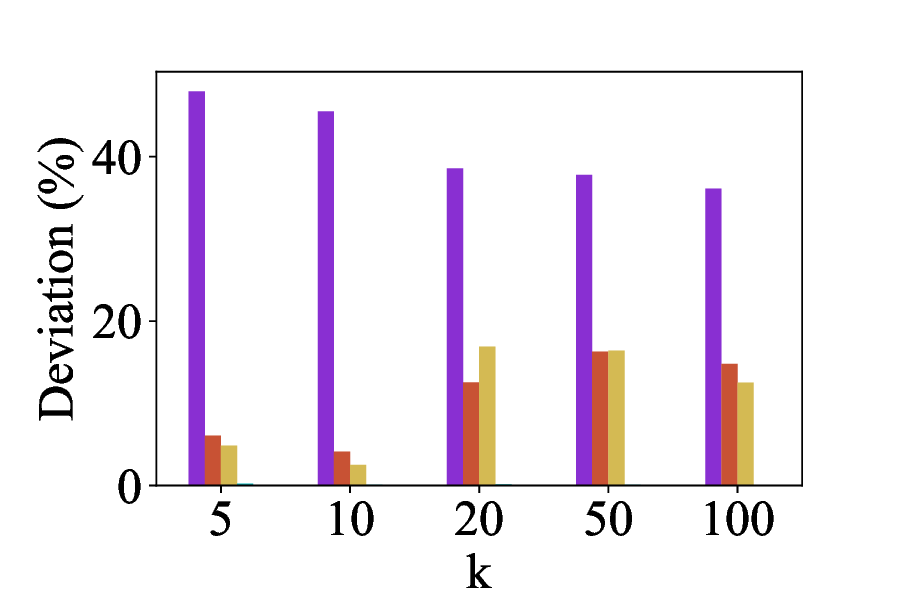}\label{subfig:var_w19_d}}
\subfloat[Twitter Dataset]{\includegraphics[trim=0.12cm 0.12cm 0.12cm 0.12cm, clip, width=0.25\textwidth]{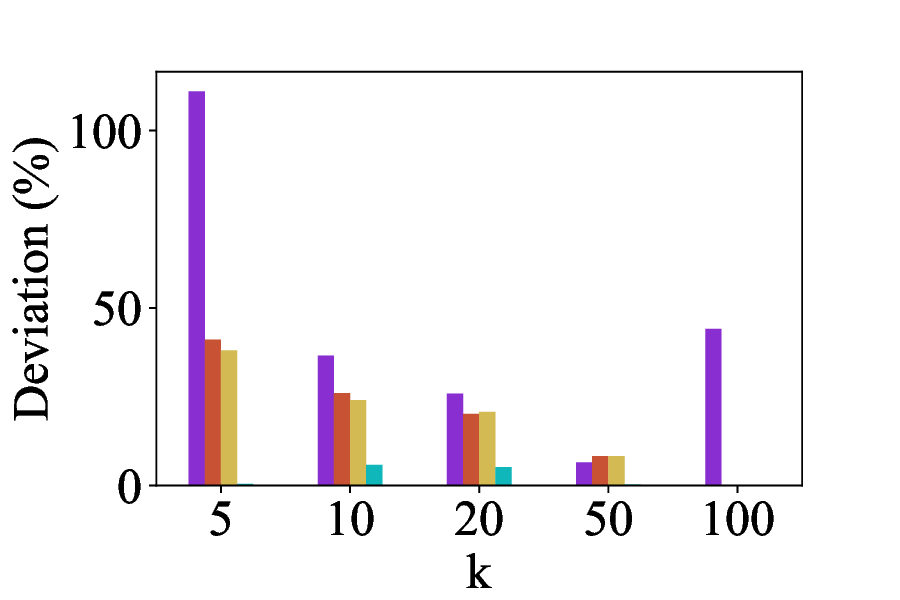}\label{subfig:var_w31_d}}
\caption{Distance deviation of $k$-th NN in B$k$Q results for varying number of query keywords $\mathcal{T}$.}
\label{fig:varying_w1_d}
\end{figure*}

\begin{figure*}[!h]
\centering
\subfloat{\includegraphics[width=\linewidth]{Figures/legend4.pdf}}
\\
\vspace{-10pt}
\setcounter{subfigure}{0}
\subfloat[Foursquare Dataset]{\includegraphics[trim=0.12cm 0.12cm 0.12cm 0.12cm, clip, width=0.25\textwidth]{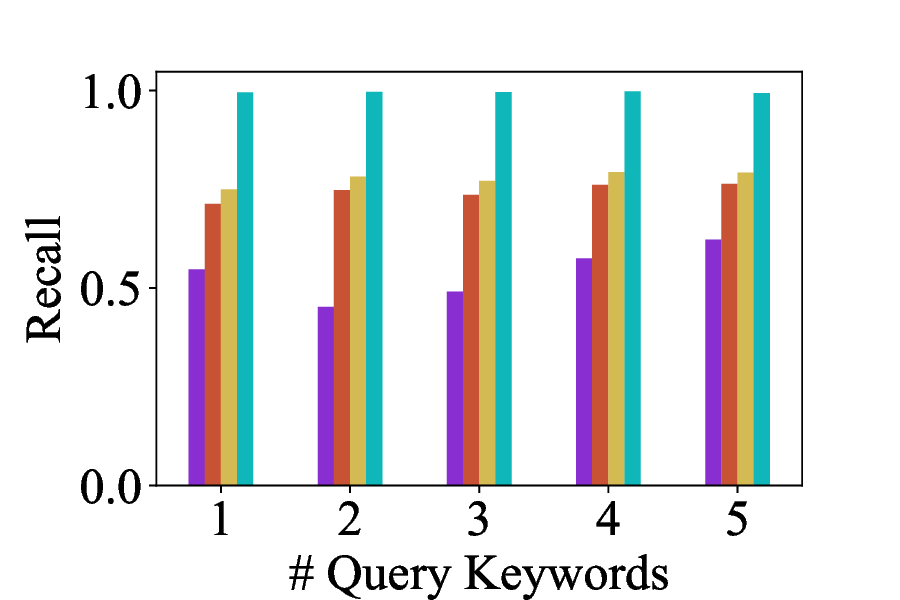}\label{subfig:var_w11_r}}
\subfloat[OSM USA Dataset]{\includegraphics[trim=0.12cm 0.12cm 0.12cm 0.12cm, clip, width=0.25\textwidth]{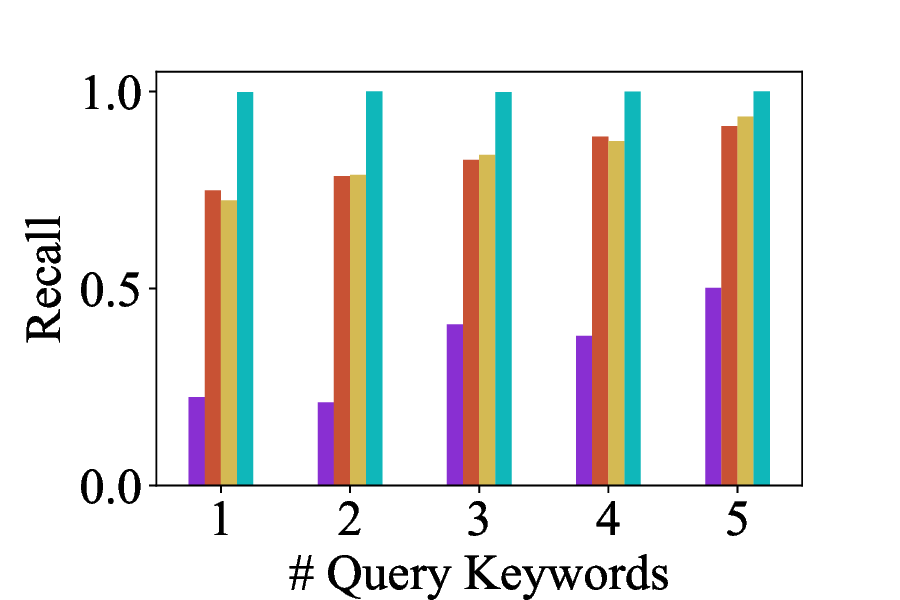}\label{subfig:var_w41_r}}
\subfloat[OSM Europe Dataset]{\includegraphics[trim=0.12cm 0.12cm 0.12cm 0.12cm, clip, width=0.25\textwidth]{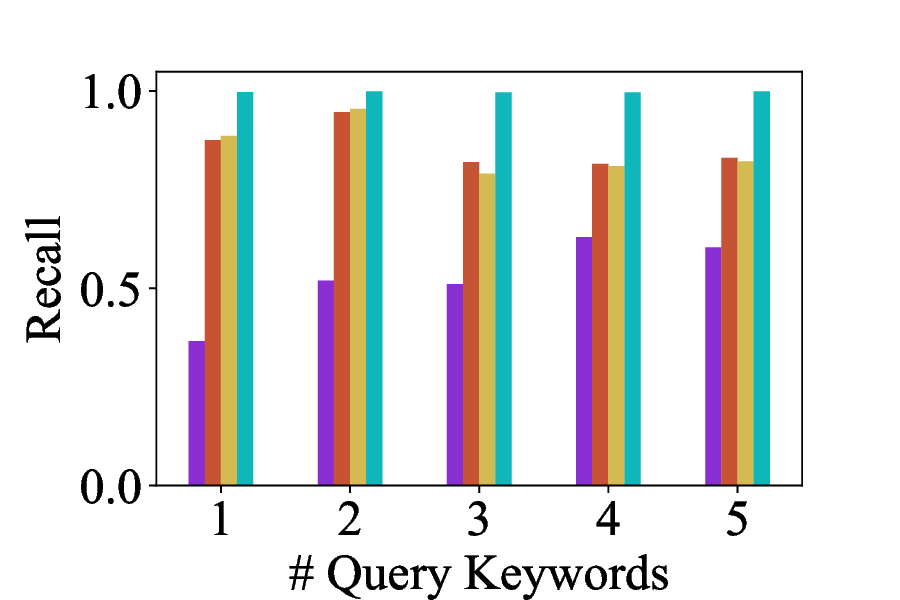}\label{subfig:var_w19_r}}
\subfloat[Twitter Dataset]{\includegraphics[trim=0.12cm 0.12cm 0.12cm 0.12cm, clip, width=0.25\textwidth]{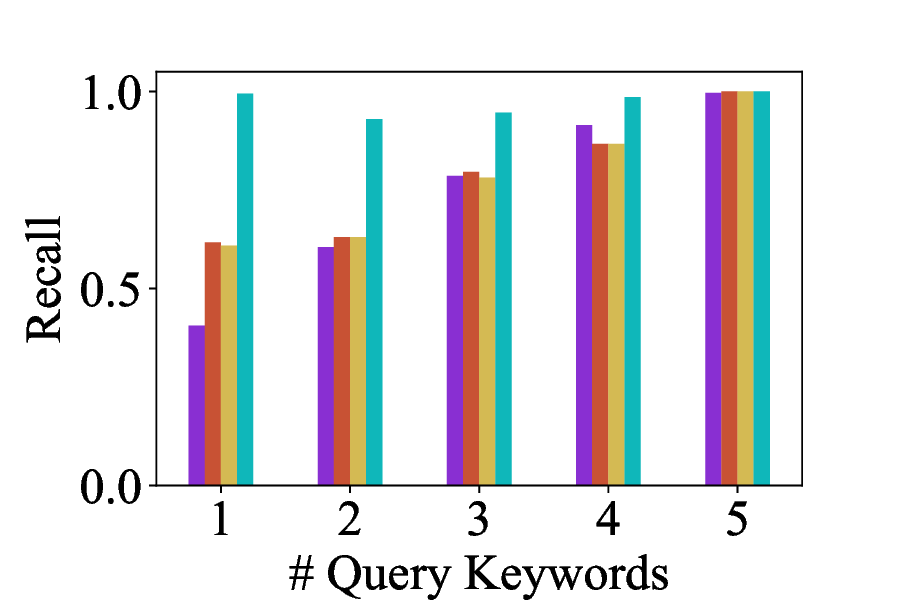}\label{subfig:var_w31_r}}
\caption{Recall of B$k$Q results for varying number of query keywords $\mathcal{T}$.}
\label{fig:varying_w1_r}
\end{figure*}

Figure~\ref{fig:varying_w1_d} shows the corresponding distance deviation of the retrieved $k$-th nearest neighbor from the exact one. As before, RSMI-BM-IR$^2$ manages to achieve a deviation close to 0\%. RSMI-BM and RSMI-BM$^*$ have a similar behavior in most cases, except for the Foursquare dataset, where  RSMI-BM$^*$ yields slightly better  results. The most interesting findings concern the Twitter dataset, and are also confirmed by the corresponding recall results in Figure~\ref{fig:varying_w1_r}. Observe that,  increasing the number of query keywords tends to improve the recall (and reduce the deviation as shown in Figure~\ref{subfig:var_w31_d}) in all RSMI-based indices over Twitter data. This is possibly because fewer objects have many keywords, which can result in less false positives. The only differentiation is RSMI-BM-IR$^2$, yielding recall values very close to 1 in all cases. Since RSMI-based processing of B$k$Q employs windows (i.e., BWQ) of side length $\ell$ (instead of circles of radius $\ell$), there is an interesting observation. The result set may contain objects that may be at a distance up to $\frac{\ell\sqrt{2}}{2}$ from the query point (close to a corner of the window). However, there may still exist objects with a smaller distance that may be outside of the applied window; these will be missed. This explains the lower recall rates for B$k$Q queries using RSMI-based schemes.


\section{Conclusions}
\label{sec:conclusions}

In this paper, we examined the case of answering Boolean spatio-textual queries by extending RSMI, a state-of-the-art spatial learned index. We presented three different indexing schemes: a loosely coupled RSMI with inverted indices, a tightly coupled scheme that augments RSMI nodes with keyword bitmaps, and a hybrid approach that attaches a traditional state-of-the-art spatio-textual index under each leaf node of RSMI. We also examined an alternative partitioning strategy that takes the textual part of the objects under consideration. Our extensive experimental evaluation indicated that, in general, the hybrid approach is faster and yields very accurate results. In the future, we plan to investigate ways for integrating the textual information in the learning process of the indices, to further improve query response times, as well as to support index updates.


\section*{Acknowledgments}
This work was supported by the EU H2020 project SmartDataLake (825041) and the EU H2020 project OpertusMundi (870228).

\bibliographystyle{abbrv}
\bibliography{References}

\end{document}